\documentclass[reprint,superscriptaddress,amsmath,amssymb,aps]{revtex4-2}

\let\hide\iffalse

\usepackage{graphicx}
\usepackage{dcolumn}
\usepackage{bm}
\usepackage{soul}
\usepackage{hyperref}
\hypersetup{pdfnewwindow=true, colorlinks=true, linkcolor=blue, anchorcolor=blue, citecolor=blue, filecolor=blue, menucolor=blue, urlcolor=blue}

\usepackage[scaled]{helvet}
\usepackage[T1]{fontenc}
\usepackage{graphicx}
\usepackage{cancel}
\usepackage{amsmath}
\usepackage{bm}
\usepackage{color}
\usepackage{braket}
\usepackage{subfiles}
\usepackage{siunitx}
\usepackage{pgfplots,pgfplotstable}
\pgfplotsset{compat=newest}
\usepgfplotslibrary{colorbrewer}
\usepgfplotslibrary{patchplots}
\usepgfplotslibrary{fillbetween}
\usetikzlibrary{bending}
\usetikzlibrary{calc,quotes}
\usetikzlibrary{arrows.meta}
\usepackage{upgreek}
\usepackage{siunitx}

\def\ve{\varepsilon}
\def\bk{{\bf k}}
\def\bq{{\bf q}}
\def\d{\delta}
\def\w{\omega}

\newcommand{\wmk}{\,\si{\watt\per\meter\per\kelvin}}
\newcommand{\mum}{$\!\!\upmu$m}
\newcommand{\ke}{$\kappa_\text{el}$}
\newcommand{\kl}{$\kappa_\text{ph}$}

\begin{document}
\def\utoden{Oden Institute for Computational Engineering and Sciences, The University of Texas at Austin, Austin, Texas 78712, USA}
\def\utphysics{Department of Physics, The University of Texas at Austin, Austin, Texas 78712, USA}
\def\utmaterial{Texas Materials Institute, The University of Texas at Austin, Austin, Texas 78712, USA}
\def\utmech{Walker Department of Mechanical Engineering, The University of Texas at Austin, Austin, Texas 78712, USA}

\title{Balanced electron and phonon heat transport in metallic $\texorpdfstring{\bm{\varepsilon}}{\varepsilon}$-TaN}

\author{Sungyeb Jung}
\affiliation{\utoden}
\affiliation{\utphysics}

\author{Hongze Li}
\affiliation{\utmaterial}
\affiliation{\utmech}

\author{Yudan Li}
\affiliation{\utmech}

\author{Noah Rossignol}
\affiliation{\utmaterial}
\affiliation{\utmech}

\author{Woongchul Choi}
\affiliation{\utmech}

\author{Yaguo Wang}
\affiliation{\utmaterial}
\affiliation{\utmech}

\author{Jianshi Zhou}
\affiliation{\utmaterial}
\affiliation{\utmech}

\author{Li Shi}
\affiliation{\utmaterial}
\affiliation{\utmech}

\author{Feliciano Giustino}
\email{fgiustino@oden.utexas.edu}
\affiliation{\utoden}
\affiliation{\utphysics}

\date{\today}

\begin{abstract}
Most materials with high thermal conductivity belong to one of two classes: metals, where heat is carried predominantly by electrons, and insulators, where heat transport is dominated by the phonon contribution. Materials that combine substantial electronic thermal conductivity and lattice thermal conductivity are rare, because the mechanisms that favor electron transport typically suppress phonon transport, and vice versa. Here, we report the theoretical prediction and experimental realization of such a material, metallic $\ve$-TaN. Our calculations predict a total thermal conductivity at room-temperature of 273~$\pm$~5\wmk\ in single crystals and 145~$\pm$~5\wmk\ in polycrystals with 0.5 \,\mum\ grains, with an unusually large lattice contribution (79\%) for a metal. The latter value is in agreement with our local transient thermoreflectance measurements on polycrystalline samples yielding $\sim$130\wmk. We show that the balanced electronic and lattice thermal conductivities of $\ve$-TaN originate from a combination of large Fermi velocity and small Fermi density of states on the electron side, and large speed of sound and wide phonon gap on the lattice side.
\end{abstract}

\maketitle

\section{Introduction}


Materials with high thermal conductivity typically fall into two classes: metals with large electronic thermal conductivity (\ke), and insulators with large lattice thermal conductivity (\kl). In metals, heat transport is carried by electrons and is primarily limited by electron-phonon scattering. Typical examples of materials in this class are Ag, Cu, and Al~\cite{Berman1975,Onn1992,Olson1993}. In insulators, on the other hand, heat is carried by phonons; in this case, \kl\ is primarily limited by phonon-phonon scattering. The material in this class with the highest known thermal conductivity is diamond~\cite{Wei1993,Ward2009}.

Beyond this standard classification, new paradigms have recently emerged with the discoveries of BAs~\cite{Lindsay2013} and $\theta$-TaN~\cite{Kundu2021}. BAs is an insulator that achieves ultra-high lattice thermal conductivity by combining closed bunched acoustic branches with a large gap between acoustic and optical phonons (AO gap); both features enable long-lived phonons by suppressing phonon-phonon scattering~\cite{Tian2018,Dames2018,Kang2018,Li2018,Lindsay2008, Lindsay2013}. $\theta$-TaN is a semimetal that achieves ultra-high thermal conductivity by combining a large AO gap with a low electronic density of states at the Fermi level~\cite{Kundu2021,Kundu2024,Kuge2022,Lee2023,Liu2023,Li2026}. In this case, electron-phonon scattering does not significantly reduce phonon lifetimes~\cite{Kundu2021,Kundu2024}. Despite its semimetallic character, $\theta$-TaN achieves high thermal conductivity almost entirely through its lattice contribution, while the electronic contribution is negligible.

\begin{figure}[t]
\centering
\begin{tikzpicture}[scale=1.0,every node/.style={scale=1.0}]
  \def\offset{0}
  \node[inner sep=0pt] at (0,0){\includegraphics[width=0.45\textwidth]{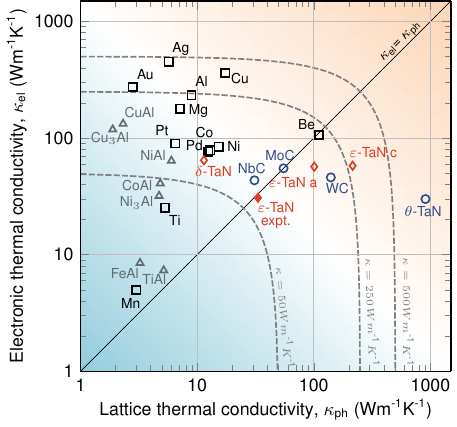}};
\end{tikzpicture}
\caption{Two-dimensional map of measured electronic and lattice thermal conductivities at 300 K for a variety of elemental metals (squares), binary metals (triangles), and calculated conductivities for semimetals with a large AO gap (circles), from Refs.~\citenum{Tong2019,Kundu2020,Chen2024,Kundu2021}. The red open symbols are the present calculations from Sec.~\ref{sec:disc}, and the filled red diamond is our measurement from Sec.~\ref{sec:transp-meas}. Dashed lines indicate total thermal conductivity isovalues. The highest theoretical value corresponds to $\kappa_\text{ph}=215\pm 5$\wmk\ and $\kappa_\text{el}=58\pm 5$\wmk for episilon-TaN.}
\label{f:map}
\end{figure}

In all these cases, heat is primarily transported either by electrons or by phonons. More generally, good thermal conductors tend to be either electron-driven or phonon-driven, as shown in Fig.~\ref{f:map}. Materials that exhibit substantial electronic thermal conductivity while retaining a large lattice thermal conductivity are rare, because the mechanisms that favor electronic heat transport typically enhance electron-phonon scattering and suppress phonon transport.

A notable example is metallic beryllium, where \ke\ and \kl\ have similar magnitude~\cite{Chen2024}. It is therefore natural to ask whether additional metallic materials can be identified in which electrons and phonons both contribute significantly to heat conduction.

Here, we establish the conditions under which metallic systems can sustain large lattice thermal conductivity without suppressing electronic heat transport. Using this understanding, we identify $\ve$-TaN as a metal with balanced electron and phonon heat transport, and we make quantitative predictions using state-of-the-art \textit{ab initio} calculations. We then confirm our predictions experimentally by synthesizing $\ve$-TaN and measuring its electronic and lattice thermal conductivities.

The paper is organized as follows. In Sec.~\ref{sec:theory}, we outline a simple scaling law with physical descriptors for both lattice and electronic thermal conductivities. Section~\ref{sec:comp} summarizes our computational setup. Section~\ref{sec:expt} describes the experimental methods, including sample preparation, structural analysis, and transport measurements. Section~\ref{sec:disc} presents the calculated and measured thermal conductivities of $\ve$-TaN, together with theoretical analyses of the origin of its high lattice thermal conductivity. In Sec.~\ref{sec:conclusion}, we summarize our findings and present our conclusions. Additional details on the derivations, convergence tests, and analyses are discussed in the Appendices, including the analysis of the related $\delta$-TaN phase.

\section{Scaling analysis}\label{sec:theory}
We begin by performing a simple scaling analysis of the electronic and lattice contributions to the thermal conductivity. To identify metals in which phonons can contribute significantly to heat transport, we focus on systems with a large AO gap as in BAs and $\theta$-TaN \cite{Lindsay2013,Kundu2021}, so that electron-phonon scattering can be regarded as an important scattering mechanism over phonon-phonon scattering.

Within the Wiedemann-Franz framework~\cite{Kittel1976}, the electronic thermal conductivity can be calculated as:
\begin{equation}\label{Eq:kel}
\kappa_\text{el} = \frac{1}{3} C_\text{el} v_\text{F}^2 \tau_\text{el},
\end{equation}
where $C_\text{el}$ is the electronic heat capacity per unit volume, $v_\text{F}$ is the Fermi velocity, 
and $\tau_\text{el}$ is the electron relaxation time. For a degenerate electron gas, the heat capacity scales as
$C_{\rm el} \sim k_\text{B}^2\, T N_\text{F}$,
where $k_\text{B}$ is the Boltzmann constant, $T$ is the temperature, and $N_\text{F}$ is the density of states at the Fermi level per unit volume (including spin degeneracy)~\cite{Kittel1976}.
In the high-temperature regime, the electron relaxation time behaves as~\cite{Giustino2017}:
\begin{equation}\label{Eq:tauel}
\frac{1}{\tau_\text{el}} \sim \frac{\Omega\, g^2 N_\text{F}}{\hbar} \frac{k_\text{B} T}{\hbar \omega}~,
\end{equation}
where $\Omega$ is the unit cell volume, $g$ is a characteristic electron-phonon matrix element,
$\omega$ is a representative phonon frequency, and numerical prefactors of order unity have been omitted. 
This relation is derived in Appendix~\ref{app:derive}.
Upon combining Eqs.~\eqref{Eq:kel} and \eqref{Eq:tauel}, we obtain the scaling law:
\begin{equation}\label{Eq:kel2}
\kappa_\text{el} \sim \frac{\hbar k_\text{B}}{\Omega} \frac{v_\text{F}^2}{g^2}\hbar\omega~.
\end{equation}
A similar reasoning can be performed for the lattice thermal conductivity. Within the relaxation time approximation, $\kappa_\text{ph}$ is given by~\cite{Ward2009}:
\begin{equation}\label{Eq:klat}
\kappa_\text{ph} = \frac{1}{3} C_\text{ph} v_\text{s}^2 \tau_\text{ph},
\end{equation}
where $C_\text{ph}$ is the lattice heat capacity per unit volume, $v_\text{s}$ is the speed of sound, and $\tau_\text{ph}$  is the phonon relaxation time.
At high temperatures, $C_\text{ph}$ approaches the Dulong-Petit limit, $C_\text{ph} \sim k_\text{B} / \Omega_\text{at}$, where $\Omega_\text{at}$ is the average volume per atom. For this simplified analysis,  we focus on the electron-phonon contribution to the phonon relaxation time. Using the standard Fermi golden rule~\cite{Giustino2017} and the free electron model, the phonon scattering rate scales as:
\begin{equation}\label{Eq:tauph}
\frac{1}{\tau_\text{ph}} \sim \frac{\Omega\, g^2 N_{\rm F}}{\hbar} \frac{v_\text{s}}{v_\text{F}}~,
\end{equation}
which has a form similar to that of the electronic rate in Eq.~\eqref{Eq:tauel}; this relation is also derived in the Appendix~\ref{app:derive}.
By combining Eqs.~\eqref{Eq:klat} and \eqref{Eq:tauph}, we obtain the scaling law:
\begin{equation}\label{Eq:klat2}
\kappa_\text{ph} \lesssim \frac{\hbar k_\text{B}}{\Omega }  
\frac{v_\text{F}^2}{g^2}N_{\rm F,at}^{-1}\frac{v_\text{s}}{v_\text{F}},
\end{equation}
where $N_{\rm F,at} = \Omega_\text{at} N_{\rm F}$ is the density of states per atom. Since in Eq.~\eqref{Eq:tauph} we do not include phonon-phonon scattering, this expression constitutes an upper bound on $\kappa_\text{ph}$ (hence the $\lesssim$ sign).

Eq.~\eqref{Eq:klat2} is similar in form to Eq.~\eqref{Eq:kel2}, except that the characteristic phonon energy $\hbar\omega$ is replaced by the factor $N_{\rm F,at}^{-1}v_\text{s}/v_\text{F}$. Therefore, for the lattice thermal conductivity to become comparable to the electronic thermal conductivity, $N_{\rm F,at}^{-1}v_\text{s}/v_\text{F}$ must approach $\hbar\omega$. Overall, this scaling analysis shows that balanced electron and phonon heat transport in a metal requires, in addition to a large AO gap: (i) large $v_\text{F}^2$, to enhance $\kappa_\text{el}$; and (ii) large $N_{\rm F,at}^{-1}v_\text{s}/v_\text{F}$, to enhance $\kappa_\text{ph}$. 

\begin{table}[t]
\caption{\label{t:fv} Calculated average Fermi velocity and density of states at the Fermi level, together with the electronic descriptor of the electronic thermal conductivity, $v_\text{F}^2$, and the electronic part of the $\kappa_{\rm ph}$ descriptor, $N_{\rm F,at}^{-1}/v_\text{F}$, for selected materials. The full lattice descriptor also includes the speed of sound, as discussed in the text.\vspace{3pt}
}
\begin{ruledtabular}
\begin{tabular}{ldddd}
\\[-8pt]
\textrm{\ }&\multicolumn{1}{c}{\textrm{$v_{\rm F}$}}&\multicolumn{1}{c}{\textrm{$N_{\rm F,at}$}}&\multicolumn{1}{c}{\textrm{$v_{\rm F}^{\rm 2}$}}&\multicolumn{1}{c}{\textrm{$N_{\rm F,at}^{-1}/v_\text{F}$}}\\
\textrm{\ }&\multicolumn{1}{c}{\textrm{(10$^{\rm 6}$m/s)}}&\multicolumn{1}{c}{\textrm{(10$^{\rm -1}$eV$^{\rm -1}$)}}&\multicolumn{1}{c}{\textrm{(10$^{\rm 12}$m$^{\rm 2}$/s$^{\rm 2}$)}}&\multicolumn{1}{c}{\textrm{(10$^{\rm -6}$ eV\,s/m)}}\\[4pt]
\colrule\\[-8pt]
Cu & 1.1 & 2.9 & 1.2 & 3.1 \\
Ag & 1.5 & 2.6 & 2.2 & 2.7 \\
Au & 1.4 & 2.8 & 1.8 & 2.6 \\
Be & 1.0 & 0.6 & 1.0 & 15.3 \\
$\theta$-TaN & 0.3 & 0.4 & 0.1 & 91.7 \\
$\ve$-TaN & 0.7 & 0.8 & 0.5 & 17.4 \\
$\delta$-TaN & 1.1 & 3.7 & 1.1 & 2.5 \\
\end{tabular}
\end{ruledtabular}
\end{table}

In Tab.~\ref{t:fv}, we inspect these trends for a few metallic and semimetallic compounds, focusing on the electronic descriptors for simplicity. We calculated the average Fermi velocity and density of states for the elemental metals, for Be, and for three polymorphs of TaN. We include elemental metals because they have the highest electronic conductivities, Be which was recently highlighted in Ref.~\citenum{Chen2024}, and polymorphs of TaN since the $\theta$ phase was predicted~\cite{Kundu2021} and subsequently confirmed~\cite{Lee2023,Li2026,Liu2023,Liu2026} to exhibit ultrahigh lattice thermal conductivity.

\begin{figure}[b]
\centering
\begin{tikzpicture}[scale=1.0,every node/.style={scale=1.0}]
  \def\offset{0}
  \node[inner sep=0pt] at (0,0){\includegraphics[width=0.475\textwidth]{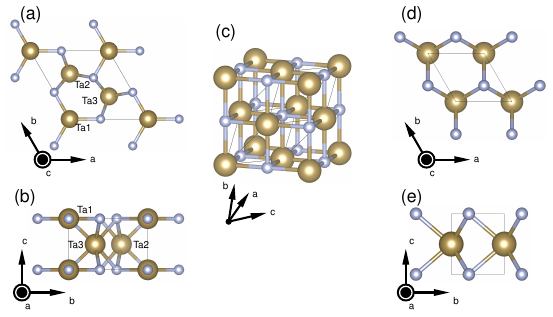}};
\end{tikzpicture}
\caption{Crystal structures of TaN polymorphs~\cite{Momma2011}. (a) and (b): Top view and side view of the crystal structure of $\ve$-TaN, respectively. Ta is in yellow, N is in gray. (c): Side view of the crystal structure of $\delta$-TaN. (d) and (e): Top and side views of the crystal structure of $\theta$-TaN, respectively. Black solid lines represent the primitive unit cells in each case.}
\label{f:struc}
\end{figure}

The elemental metals have the largest $v_\text{F}^2$, consistent with their large electronic thermal conductivities, but small $N_{\rm F,at}^{-1}/v_\text{F}$, consistent with a suppressed lattice contribution. Conversely, $\theta$-TaN has the largest value of $N_{\rm F,at}^{-1}/v_\text{F}$ but the smallest $v_\text{F}^2$, consistent with its thermal conductivity dominated by the lattice contribution~\cite{Kundu2021}. The only materials in Tab.~\ref{t:fv} combining sizable $v_\text{F}^2$ with large $N_{\rm F,at}^{-1}/v_\text{F}$ are Be and $\ve$-TaN. The case of Be was analyzed in Ref.~\citenum{Chen2024}, while $\ve$-TaN has not yet been investigated. These considerations motivate an in-depth analysis of $\ve$-TaN using \textit{ab initio} calculations and experiments, as reported in the following sections.

\section{Computational details}\label{sec:comp}
All density functional theory (DFT) calculations are performed using planewaves and pseudopotentials as implemented in the \textsc{Quantum ESPRESSO} materials simulation suite~\cite{Giannozzi2017}. We employ the scalar-relativistic local density approximation (LDA) to the DFT exchange and correlation functional~\cite{Ceperley1980,Perdew1981} and optimized norm-conserving Vanderbilt (ONCV) pseudopotentials from the \textsc{PseudoDojo} library~\cite{Hamann2013,VanSetten2018}. We set the planewaves kinetic energy cutoff to 150~Ry, and we converge the total energy with a threshold of $10^{-12}$~Ry. We calculate vibrational eigenfrequencies and eigenmodes using density functional perturbation theory (DFPT)~\cite{Baroni1987,Giannozzi1991,Gonze1997,Baroni2001}. We employ a tight convergence criterion in all phonon calculations by setting the squared relative error in the DFPT self-consistency cycle (``tr2\_ph'') to $10^{-16}$.

For completeness we also investigate the $\delta$ phase of TaN. In the harmonic approximation, this phase exhibits soft modes, therefore we include anharmonic corrections using the special displacement method~\cite{Zacharias2020,Zacharias2023} as implemented in the \textsc{EPW/ZG} code~\cite{Lee2023epw}. In this case, to generate the polymorphous structure, we employ a $2\times 2\times 2$ supercell.

For the third order and fourth order interatomic force constants, we perform finite-difference supercell calculations via \textsc{Phonopy}~\cite{togo2023,togo2023.1}, \textsc{thirdorder.py}, and \textsc{fourthorder.py}~\cite{Li2014,Han2022}. All quantities are interpolated from coarse to fine Brillouin zone grids via Fourier interpolation (phonon properties)~\cite{Baroni2001}, and Wannier-Fourier interpolation for electron bands~\cite{Marzari1997,Souza2001} and electron-phonon matrix elements~\cite{Giustino2007}, via \textsc{Wannier90}~\cite{Pizzi2020} and \textsc{EPW}~\cite{Lee2023epw}.
In Tab.~\ref{t:conv} we list all coarse grids and fine grids used in this work. All grids are uniform and unshifted. The numbers of grid points along the $ab$ plane and the $c$ axis in $\varepsilon$-TaN differ due to the anisotropic unit cell.

\begin{table}[t]
\caption{\label{t:conv} Brillouin zone grids and supercell sizes employed in this work for electrons ($\mathbf{k}$-grid) and phonons ($\mathbf{q}$-grid). The rows for $\kappa_\text{el}$ correspond to the electronic BTE solved via \textsc{EPW}. The row labeled ``EPW $\mathbf{k}$-grid'' refers to the coarse k-grid used for Wannier interpolation. The rows for $\tau_\text{ph}$ indicate the grids used to compute phonon relaxation times from electron-phonon scattering, subsequently employed in \textsc{SchengBTE}. The row marked ``IFC supercell'' indicates the Born-von K\'arm\'an supercells employed to compute interatomic force constants from finite differences using \textsc{Phonopy}, \textsc{thirdorder.py}, and \textsc{fourthorder.py}. In the case of $\delta$-TaN, harmonic phonons are soft. To obtain stable modes, we employed the anharmonic special displacement method, and in the row ``DFPT $\bf q$-grid'' we report the size of the corresponding supercell.\vspace{3pt}}
\begin{ruledtabular}
\begin{tabular}{llll}
\textrm{\ }&
\textrm{$\varepsilon$-TaN}&
\textrm{$\delta$-TaN}&
\textrm{$\theta$-TaN}\\[2pt]
\hline\\[-8pt]
DFT $\mathbf{k}$-grid & 12$\times$12$\times$20 & 20$\times$20$\times$20 & 20$\times$20$\times$20\\
DFPT $\mathbf{q}$-grid & 6$\times$6$\times$10 & 2$\times$2$\times$2 & 4$\times$4$\times$4\\
EPW $\mathbf{k}$-grid & 6$\times$6$\times$10 & 8$\times$8$\times$8 & 8$\times$8$\times$8\\[2pt]
\hline\\[-8pt]
$\kappa_{\text{el}}$ fine $\mathbf{k}$-grid & 96$\times$96$\times$160 & 120$\times$120$\times$120 & 100$\times$100$\times$100\\
$\kappa_{\text{el}}$ fine $\mathbf{q}$-grid & 96$\times$96$\times$160 & 120$\times$120$\times$120 & 100$\times$100$\times$100\\
$\tau_{\text{ph}}$ fine $\mathbf{k}$-grid & 60$\times$60$\times$100 & 96$\times$96$\times$96 & 84$\times$84$\times$84\\
$\tau_{\text{ph}}$ fine $\mathbf{q}$-grid & 12$\times$12$\times$20 & 16$\times$16$\times$16 & 16$\times$16$\times$16\\[2pt]
\hline\\[-8pt]
IFC supercell & 3$\times$3$\times$5 & 5$\times$5$\times$5 & 5$\times$5$\times$5\\
$\kappa_{\text{ph}}$ $\mathbf{q}$-grid & 12$\times$12$\times$20 & 16$\times$16$\times$16 & 16$\times$16$\times$16\\
\end{tabular}
\end{ruledtabular}
\end{table}

Third- and fourth-order interatomic force constants are computed using supercells whose size is reported in Tab.~\ref{t:conv}, and a $2\times2\times2$ Brillouin zone grid in all cases. In calculating the three-phonon contribution to the lattice thermal conductivity, we consider interactions up to the sixth-nearest neighbors for $\delta$-TaN, and up to eighth-nearest neighbors for $\varepsilon$-TaN. In the case of $\theta$-TaN, we include terms up to the sixth-nearest neighbors, as in our previous work~\cite{Lee2023}. For the four-phonon contribution, we include interactions up to second-nearest neighbors for $\varepsilon$- and $\delta$-TaN. In the case of $\theta$-TaN, we include terms up to the third-nearest neighbors. Detailed convergence tests with respect to number of neighbors are reported in Appendix~\ref{app:convtests-lattherm}.

We perform calculations of the lattice thermal conductivity using the phonon Boltzmann transport equation (BTE) as implemented in \textsc{ShengBTE}~\cite{Li2014}. In the collision integral we include three-phonon scattering~\cite{Li2014}, four-phonon scattering~\cite{Feng2016,Feng2017}, phonon-isotope scattering~\cite{Lindsay2013}, and electron-phonon scattering~\cite{Giustino2017,Liao2015}. We obtain the total scattering rate from Matthiessen's rule. 

Figure~\ref{f:struc} shows the primitive unit cells and $\varepsilon$-TaN, $\delta$-TaN, and $\theta$-TaN. The $\varepsilon$ phase consists of six atoms per cell and has space group $P\bar{6}2m$, and to a first approximation it can be understood as a distorted $\sqrt{3}\times\sqrt{3}\times 1$ supercell of the $\theta$ phase. The $\theta$ phase consists of two atoms per unit cell and has space group $P\bar{6}m2$. The $\delta$ phase has two atoms per cell and space group $Fm\bar{3}m$.
The optimized lattice parameters of $\varepsilon$-TaN are $a = b = 5.136~\si{\angstrom}$ and $c = 2.868~\si{\angstrom}$, respectively. The optimized lattice parameter of $\delta$-TaN is $a=  4.337~\si{\angstrom}$. The optimized lattice parameters of $\theta$-TaN are $a = b =  2.900~\si{\angstrom}$ and $c = 2.844~\si{\angstrom}$, respectively.

We compute the electronic thermal conductivity using the \textit{ab initio} BTE as implemented in \textsc{EPW}~\cite{Lee2023epw}. The equations are a generalization of the standard electronic heat transport equations~\cite{Georg2006} to include \textit{ab initio} electron relaxation times from electron-phonon scattering, and will be available in \textsc{EPW} v6.2.

The calculations of Fermi velocity and density of states reported in Tab.~\ref{t:fv} for Cu, Ag, Au, and Be are performed via DFT band structure calculations. We use a planewaves cutoff of 150~Ry in all cases, and employ the tetrahedron method to evaluate the density of states. 

Appendix~\ref{app:convtests} describe initial orbital projection for Wannier interpolation and comparison between DFT and Wannier band structures. We also present Brillouin zone sampling convergence tests for $\kappa_{\mathrm{el}}$ and $\kappa_{\mathrm{ph}}$ and nearest neighbor in high-order IFCs convergence tests for $\kappa_{\mathrm{ph}}$.

\section{Experimental setup}\label{sec:expt}

\subsection{Sample preparation}
We synthesize $\varepsilon$-TaN samples by direct reaction between Ta (BeanTown Chemical, 99.95\%) and N$_2$ (Linde Gas \& Equipment, 99.999\%). The reaction was carried out in an infrared-heating image furnace (NEC SC-M35HD) at temperature $T=1673$-2073\,K and partial nitrogen pressure $p_{\text{N}_\text{2}}=0.2$-0.4\,MPa. Up to 10 wt\% of binding solvent (Ni, Co, Fe) was added to the Ta powder to facilitate the growth of $\varepsilon$-TaN grains and create low-porosity composites. Since Ta and the binding solvent are sensitive to oxygen at elevated temperatures, an oxygen trap (Restek) was used to further purify the N$_2$ gas. The diameter of the synthesized rod ranges from 4\,mm to 8\,mm, an the length reaches up to 80\,mm.

\subsection{Structural characterization}
\begin{figure}[t]
\centering
\begin{tikzpicture}[scale=1.0,every node/.style={scale=1.0}]
  \def\offset{0}
  \node[inner sep=0pt] at (0,0){\includegraphics[width=0.42\textwidth]{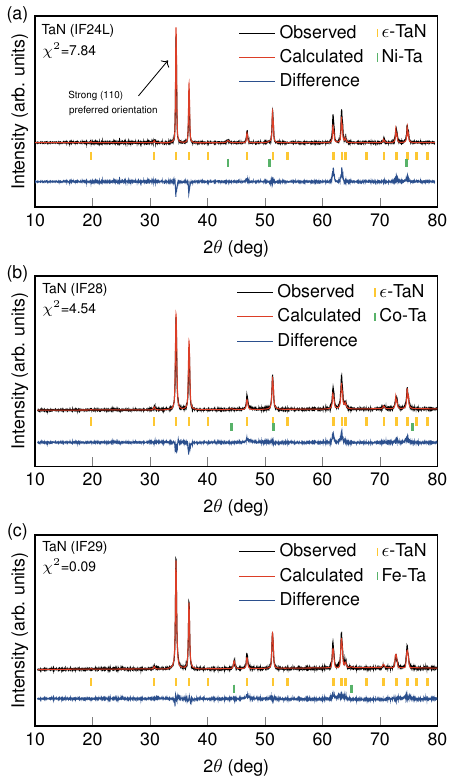}};
\end{tikzpicture}
\caption{Typical XRD patterns with Rietveld refinement results in (a) $\varepsilon$-TaN/Ni-Ta composite (sample IF24L) with $\sim$ 92~wt~\% of $\varepsilon$-TaN, (b) $\varepsilon$-TaN/Co-Ta composite (sample IF28) with $\sim$ 96~wt~\% of $\varepsilon$-TaN, and (c) $\varepsilon$-TaN/Fe-Ta composite (sample IF29) with $\sim$ 90~wt~\% of $\varepsilon$-TaN.}
\label{f:pxrd_if}
\end{figure}

The phase composition was confirmed in a polished $\varepsilon$-TaN composite specimen by X-ray diffraction (XRD) with a Rigaku Miniflex 600 Diffractometer using Cu-K$\alpha$ radiation. Rietveld refinement was performed for the XRD data using the FULLPROF program \cite{RODRIGUEZCARVAJAL199355}. Typical XRD patterns of the Ni-, Co-, and Fe-base $\varepsilon$-TaN composites are shown in Fig.~\ref{f:pxrd_if}.

\begin{figure}[t]
\centering
\begin{tikzpicture}[scale=1.0,every node/.style={scale=1.0}]
  \def\offset{0}
  \node[inner sep=0pt] at (0,0){\includegraphics[width=0.475\textwidth]{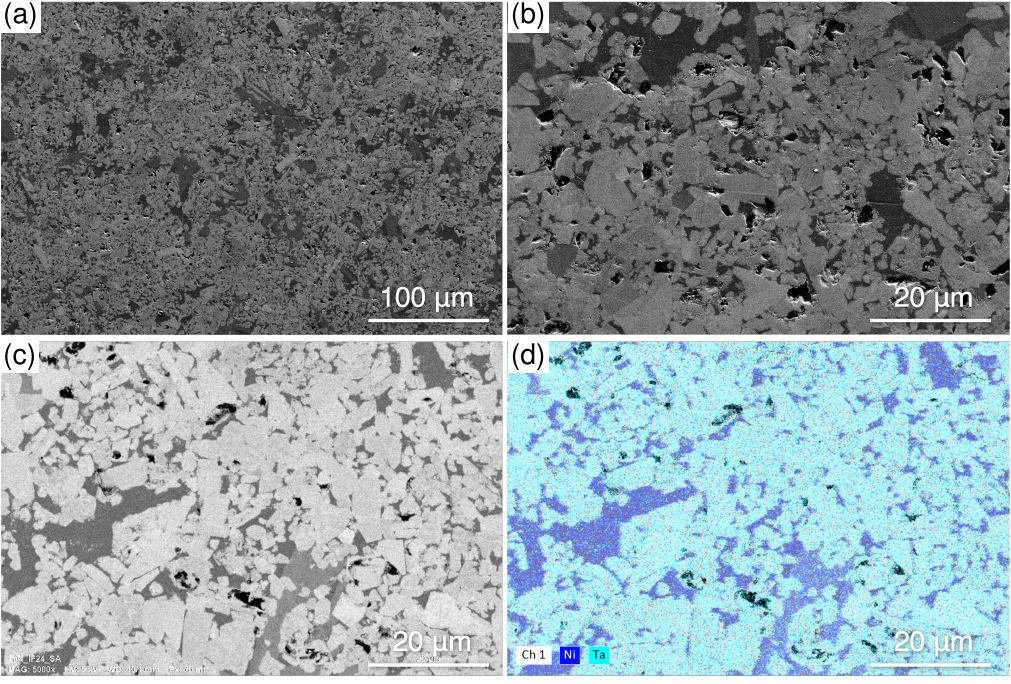}};
\end{tikzpicture}
\caption{Typical microscopic morphology of a $\varepsilon$-TaN/Ni-Ta composite (sample IF24S). Secondary electron (SE) images are shown in (a) and (b). (c) and (d) show a backscattered electron (BSE) image and a Ta-Ni elemental map collected with EDS, respectively, of the same region.}
\label{f:sem_eds_if24}
\end{figure}

Microscopic morphology and elemental distribution were investigated by scanning electron microscopy (SEM; Thermo Scientific Apreo~2) with energy-dispersive X-ray spectroscopy (EDS; Bruker XFlash 7) capability. For example, in the Ni-based $\varepsilon$-TaN composite shown in Fig.~\ref{f:sem_eds_if24}, micrometer-sized TaN grains are embedded in the Ni-Ta alloy, forming a low-porosity composite. The lighter color regions in the backscattered electron (BSE) image, Fig.~\ref{f:sem_eds_if24}(c), correspond to heavier elements or Ta-rich regions, while the darker regions are related to lighter elements or Ni-rich regions. The Ni and Ta distributions shown by the EDS elemental mapping, Fig.~\ref{f:sem_eds_if24}(d), are consistent with the BSE image.

\begin{figure}[t]
\centering
\begin{tikzpicture}[scale=1.0,every node/.style={scale=1.0}]
  \def\offset{0}
  \node[inner sep=0pt] at (0,0){\includegraphics[width=0.475\textwidth]{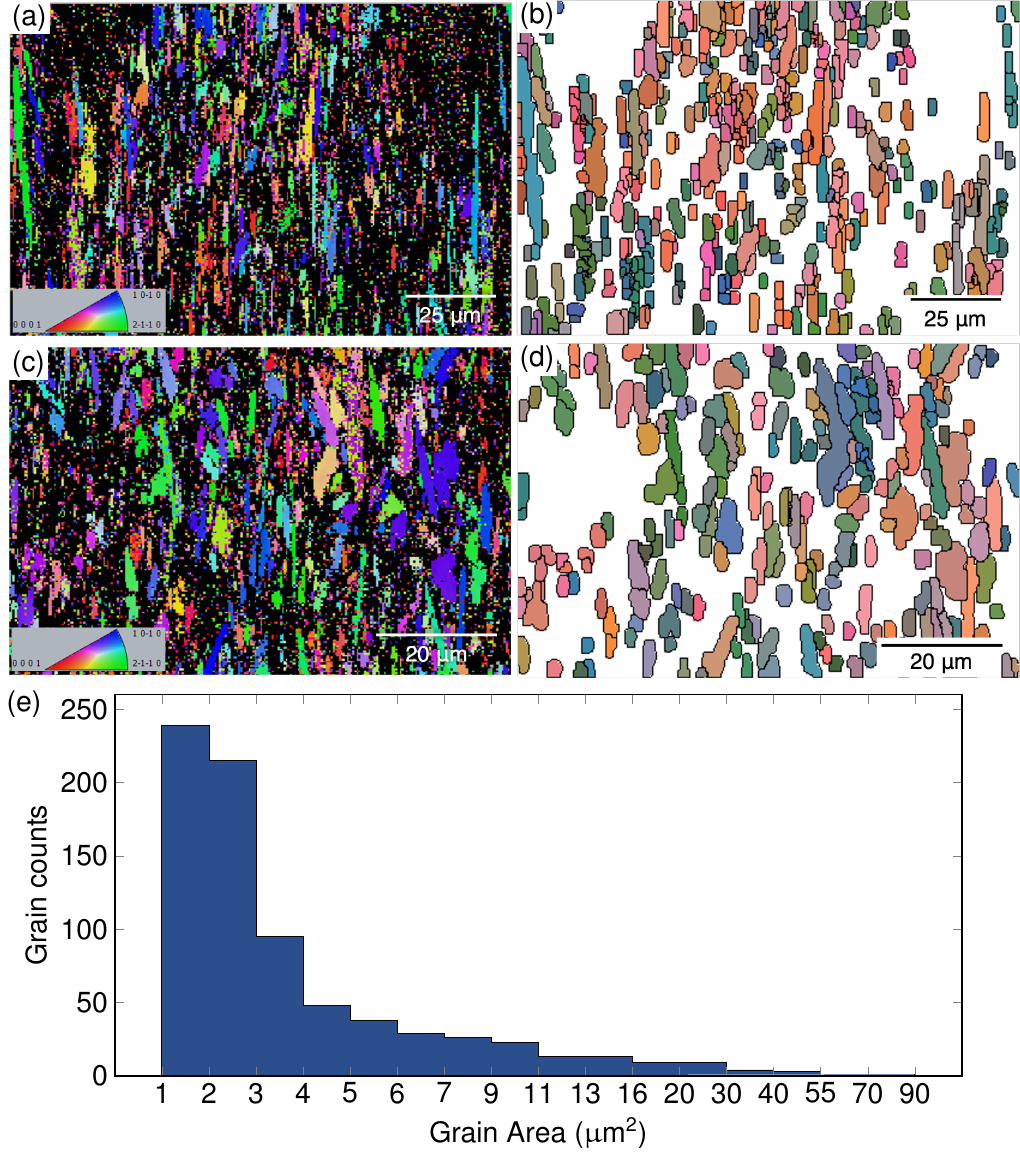}};
\end{tikzpicture}
\caption{Typical EBSD mapping of a $\varepsilon$-TaN/Ni-Ta composite (sample IF24L). (a) and (c) show the grain orientation map within two regions of interest, respectively. (b) and (d) show the corresponding identified grains with an area larger than 1 $\upmu$m$^2$ and an orientation tolerance of 5$^\circ$. The orientations were indexed using the $\varepsilon$-TaN structure model ($P6/mmm$) and color coded. (e) Grain size distribution of the indexed grains within both regions of interest.}
\label{f:ebsd_if24}
\end{figure}

Electron backscatter diffraction (EBSD; EBSD-LUMIS in Thermo Scientific Scios 2 DualBeam System) was used to examine the grain structure and grain size distribution. The polished surfaces for the EBSD analysis were prepared with diamond lapping films (30, 20, 9, 3, 1, 0.5-$\upmu$m grade), followed by final polishing with 0.02\,$\upmu$m colloidal silica. Single spot probe measurements were performed to evaluate pattern quality with conditions of $1\times1$ binning (i.e., no binning), 400-500\,ms exposure, and a gain of 0-2. EBSD mapping was performed in the region of interest (ROI) with conditions of $2\times2$ binning, 15-20\,ms exposure, and a gain of 15-20. During mapping, the $P6/mmm$ structure model of $\varepsilon$-TaN was used to index the Kikuchi patterns collected at each point by Hough indexing. For example, on the Ni-based $\varepsilon$-TaN composite (sample labeled ``IF24L''), the indexation determines the grain orientation at each point and generates the orientation maps shown in Figs.~\ref{f:ebsd_if24}(a) and \ref{f:ebsd_if24}(c). Counting grains with an area larger than 1\,$\upmu$m$^2$ and a misorientation tolerance of 5$^\circ$, a total of 765 grains were identified within three ROIs, as shown in Figs.~\ref{f:ebsd_if24}(b) and \ref{f:ebsd_if24}(d). The corresponding grain size distribution is shown in Fig.~\ref{f:ebsd_if24}(e). Due to the spatial resolution limit of the EBSD mapping, grains with an area of approximately 1~$\mu$m$^2$ appear most frequently in Fig.~\ref{f:ebsd_if24}(e). However, many smaller grains are expected to be present but are not resolved by the EBSD mapping. Thus, the peak frequency is expected to occur for a grain size below 1~$\mu$m$^2$. In addition, each grain in the EBSD map can still contain multiple coherent crystallites with a similar crystal orientation and low-angle crystalline boundaries including twin defects and dislocations. EBSD analysis was also performed on Co- and Fe-based $\varepsilon$-TaN composites, which are included in Tab.~\ref{t:samplelist} for four samples labeled as ``IF24S'', ``IF24L'', ``IF28'', and ``IF29'', respectively. Qualitatively similar results are obtained on the four samples, although sample IF24L appears to have the largest grains among them.

\begin{table}[b]
\caption{\label{t:samplelist} Measured Phase composition for three different $\varepsilon$-TaN composites. Phase composition is determined through the Rietveld refinement of the XRD patterns.\vspace{3pt}}
\begin{ruledtabular}
{\renewcommand{\arraystretch}{1.2}
\begin{tabular}{lc}
\textrm{Sample}&
\textrm{Phase composition}\\[2pt]
\colrule\\[-11pt]
IF24S & $\sim70$ wt\% $\varepsilon$-TaN and $\sim30$ wt\% Ni-Ta \\
IF24L & $\sim92$ wt\% $\varepsilon$-TaN and $\sim\phantom{0}8$ wt\% Ni-Ta \\
IF28 & $\sim96$ wt\% $\varepsilon$-TaN and $\sim\phantom{0}4$ wt\% Co-Ta \\
IF29 & $\sim90$ wt\% $\varepsilon$-TaN and $\sim10$ wt\% Fe-Ta \\
\end{tabular}
}
\end{ruledtabular}
\end{table}

\subsection{Characterization of transport properties}\label{sec:transp-meas}

Measurements of the bulk average thermal conductivity $\kappa$ were performed using laser flash analysis (LFA; NETZSCH LFA 457), differential thin film resistance thermometry (RTh)~\cite{10.1063/5.0061049}, and thermal transport option (TTO) in a physical property measurement system (PPMS; Quantum Design)~\cite{Maldonado1992}. Measurements of the local thermal conductivity of $\ve$-TaN domains and binder regions were performed using a picosecond transient thermoreflectance (psTTR) system \cite{Jeong03072019, Ye02102023}.
These complementary measurements were used to separate the contribution of $\varepsilon$-TaN domains from that of intergranular Ni-Ta alloy phases. We note that $\ve$-TaN domains still consist of smaller TaN bounded by large-angle grain boundaries, which contain even smaller coherent crystallites and low-angle crystallite boundaries.

\begin{figure}[t]
\centering
\begin{tikzpicture}[scale=1.0,every node/.style={scale=1.0}]
  \def\offset{0}
  \node[inner sep=0pt] at (0,0){\includegraphics[width=0.475\textwidth]{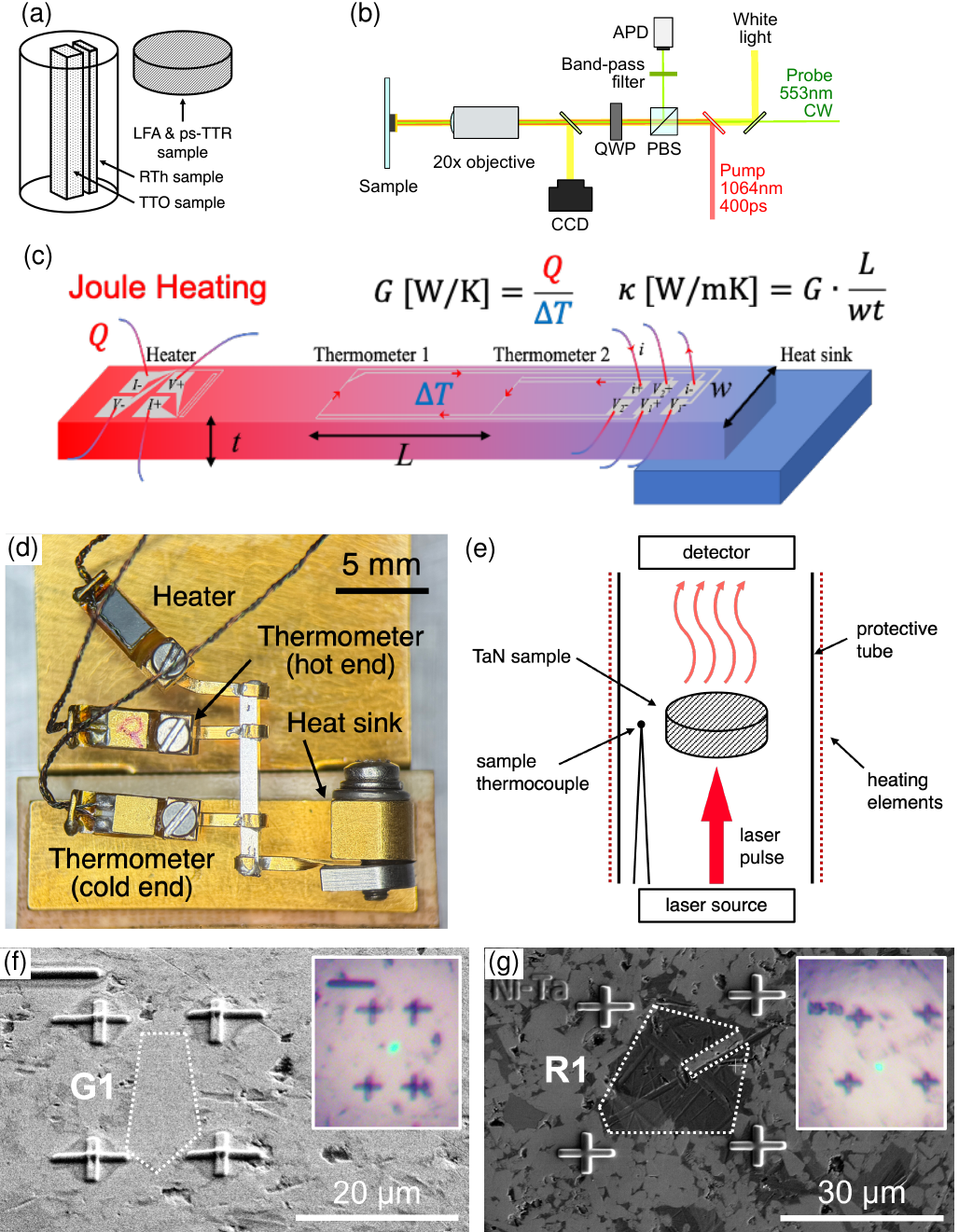}};
\end{tikzpicture}
\caption{Thermal measurements setup. (a) Schematic, not to scale, of the LFA, RTh, TTO, and psTTR samples. (b) Optical layouts of the psTTR from \cite{Ye02102023}, in which APD stands for avalanche photodiode detector and QWP stands for quarter-wave plate. (c) Schematic, not to scale, of the RTh setup. (d) Photograph of fabricated device for the TTO measurement. (e) Schematic, not to scale, of the LFA setup. Dash lines in the SEM images mark (f) an $\varepsilon$-TaN domain} (G1) and (g) a Ni-Ta alloy region (R1) in sample IF24S. The insets in (f) and (g) show the optical view of the psTTR probe beam collected by a charge-coupled device (CCD).
\label{f:measurement_setup}
\end{figure}

LFA samples are disks cut from composite rods, with a diameter ranging from 5.5 to 5.9\,mm and a thickness greater than 1\,mm, as shown in Fig.~\ref{f:measurement_setup}(a). The cross section of the LFA disks is perpendicular to the axial direction of the composite rods. The heat capacity $C_\text{p}$ of the LFA disks is measured by differential scanning calorimetry (DSC). As shown in Fig.~\ref{f:measurement_setup}(e), by applying a laser pulse and collecting the resulting thermal response, LFA measures the thermal diffusivity ($\alpha$) of samples from 300 to 580\,K, and the corresponding $\kappa$ is calculated through $\kappa=\alpha\rho_\text{m} C_\text{p}$, where $\rho_\text{m}$ is mass density.

The RTh samples are rectangular plates with a typical dimension of $6.0\times1.0\times0.4$ mm$^3$, as shown in Fig.~\ref{f:measurement_setup}(a). The longest dimension of the RTh plates is along the growing direction of the composite rods. The RTh method is based on the deposition and patterning of thin film resistive heaters and thermometers directly on the sample surface. A 800-nm silicon nitride (SiN$_x$) dielectric layer is deposited on the surface for electrical insulation. Both the heater and thermometers were patterned from 80\,nm Au/Cr metal deposited on the sample surface using electron beam lithography (EBL). The end of the thermometer side of the prepared TaN sample is joined with silver epoxy to a copper heat sink in the chip carrier, while the serpentine resistive heater is patterned on the suspended end. The same sensing current travels through the two thermometers in series, and lock-in amplifiers are used to measure the resistance of each thermometer independently. The temperature at each thermometer position is obtained from the measured resistance change and the temperature coefficient of resistance (TCR), and the temperature drop ($\delta\theta$) between the two thermometers is used with a 1D heat conduction model to obtain the thermal conductivity of the sample. Heat loss to the bonding wires of the heater is obtained directly from two measurements with different numbers of lead wires attached to the heater.

The TTO samples are bars with a typical dimension of $0.8\times0.8\times8.0$ mm$^3$, as shown in Fig.~\ref{f:measurement_setup}(a). The longest dimension of the TTO bars is along the growing direction of the composite rods. The TTO uses the one-heater-two-thermometer setup as shown in Fig.~\ref{f:measurement_setup}(d), where the heater applies heat pulses and the thermometers measure the thermal relaxation. The TTO is used to measure $\kappa$ from 3 to 390\,K. The thermal conductivity at 300\,K for all measured samples is reported in Tab.~\ref{t:samplelist}. Electrical resistivity ($\rho$) and Seebeck coefficient ($S$) were measured in the same samples simultaneously with $\kappa$ from 3\,K to 390\,K using the TTO. This allows us to correlate the measured $\kappa$ and $\rho$ to understand the electronic contribution to the total thermal conductivity of $\varepsilon$-TaN. The experimental data included in Fig.~\ref{f:map} is based on the TTO measurement.

The psTTR measurements were taken using the setup shown in Fig.~\ref{f:measurement_setup}(b). A 105\,nm Au layer is deposited on the surface of the $\varepsilon$-TaN composite sample prepared following the same procedure as for the EBSD samples. The sample surface is rapidly heated by the pump, a 1064\,nm Q-switched laser with a 400\,ps pulse width (Bright Solutions Wedge XF), and the temperature change is monitored by measuring the intensity of the probe, a reflected continuous-wave (CW) 553\,nm diode pumped solid state laser (Oxxius LaserBoxx). The pump spot size is $\sim$25\,$\upmu$m, and the probe spot size is $\sim$3\,$\upmu$m. A finite difference method (FDM) simulation is used to predict the transient temperature change of the sample surface with a multi-layer 1D thermal diffusion model; the thermal conductivity $\kappa$ is extracted by fitting the transient cooling of the surface after the pulse arrives. 

We performed a series of psTTR measurements on the same $\varepsilon$-TaN/Ni-Ta sample, labeled IF24S, to assess the spatial variation and reproducibility of the measured thermal conductivity. Repeated measurements were conducted at multiple locations within a $\varepsilon$-TaN particle; see Fig.~\ref{f:measurement_setup}(f), across several neighboring $\varepsilon$-TaN domains, and within a Ni-Ta alloy region; see Fig.~\ref{f:measurement_setup}(g).

\section{Results and discussion}\label{sec:disc}
Figure~\ref{f:map} shows the calculated room-temperature $\kappa_\text{el}$ and $\kappa_\text{ph}$ of $\ve$-TaN together with literature data for elemental metals, binary metals, Be, $\theta$-TaN, and related semimetals~\cite{Tong2019,Kundu2020,Chen2024,Kundu2021}. Unlike conventional metals, which cluster in the region of electron-dominated heat transport, and semimetallic $\theta$-TaN, which lies in the phonon-dominated region, $\ve$-TaN occupies an intermediate regime in which both electrons and phonons contribute substantially. 

Our calculations predict a room-temperature thermal conductivity as high as $273\pm 5$~\wmk\ in single crystals, with $\kappa_\text{ph}=215\pm 5$\wmk\ and $\kappa_\text{el}=58\pm 5$\wmk\ along the $c$ axis. This thermal conductivity exceeds the 237~\wmk\ value of Al, and places $\ve$-TaN among metallic systems with unusually high lattice thermal conductivity.

\begin{figure}[t]
\centering
\begin{tikzpicture}[scale=1.0,every node/.style={scale=1.0}]
  \def\offset{0}
  \node[inner sep=0pt] at (0,0){\includegraphics[width=0.45\textwidth]{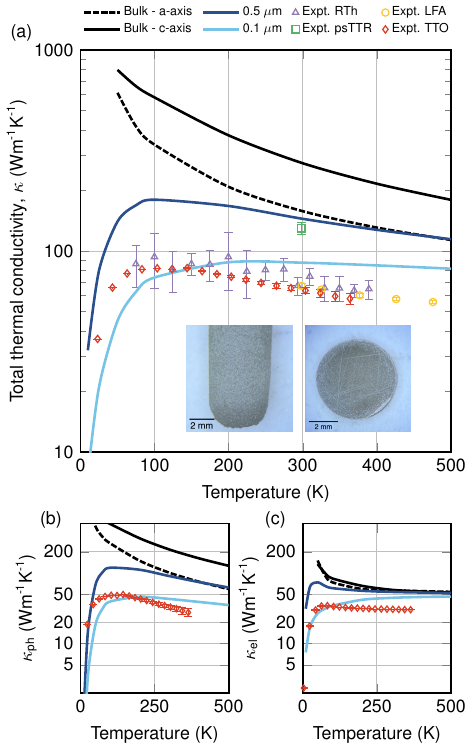}};
\end{tikzpicture}
\caption{(a) Total thermal conductivity of $\ve$-TaN. The solid and dashed black lines are our calculations for bulk $\ve$-TaN along the $c$-axis and $a$-axis, respectively. The dark blue and light blue lines are calculations including grain-boundary scattering, for 0.5~\mum\  and 0.1~\mum\ grains, respectively. Symbols are our experimental data obtained by RTh (triangles), psTTR (square), LFA (circles), and TTO (diamonds). The insets are photographs of the IF24S sample; see Tab.~\ref{t:samplelist}. (b) and (c) Decomposition of the total thermal conductivity of (a) into electronic and lattice contributions, respectively. The color code is the same as in (a).}
\label{f:therm}
\end{figure}

In Fig.~\ref{f:therm}(a) we compare our measured $\kappa$ with our \textit{ab initio} calculations. Symbols are our average bulk measurements obtained via TTO (red diamonds), LFA (yellow circles), and RTh (purple triangles), as well as a local psTTR (green square) measurement. The room-temperature kappa values measured by LFA, RTh, and TTO are all in the range between 60~\wmk\ and 70~\wmk\ in $\varepsilon$-TaN/Ni-Ta sample (IF24S). The full set of measurements for all characterized samples together with the corresponding uncertainty (of the order of $\pm 5-10$\wmk) are reported in Appendix~\ref{app:therm}.

Our measured $\kappa$ is comparable to but lower than our theoretical prediction (black lines), with the highest conductivity of 130~\wmk\ obtained for the local psTTR measurement. To perform a meaningful comparison between theory and experiment, we evaluate the isotropically averaged thermal conductivity and we include grain-boundary scattering. The results are shown in Fig.~\ref{f:therm}(a) as blue and cyan lines for 0.5~\mum\ and 0.1~\mum\ grains, respectively, and are closer to experimental results than the bulk calculation data. In particular, the calculation for 0.5~\mum\ grains matches our psTTR on a local domain. The lower bulk average $\kappa$ measured via LFA, TTO, and RTh can be explained in terms of the Ni-Ta alloy present between the grains, which has a relatively low room-temperature thermal conductivity of 15~\wmk. 

\begin{figure}[t]
\centering
\begin{tikzpicture}[scale=1.0,every node/.style={scale=1.0}]
  \def\offset{0}
  \node[inner sep=0pt] at (0,0){\includegraphics[width=0.475\textwidth]{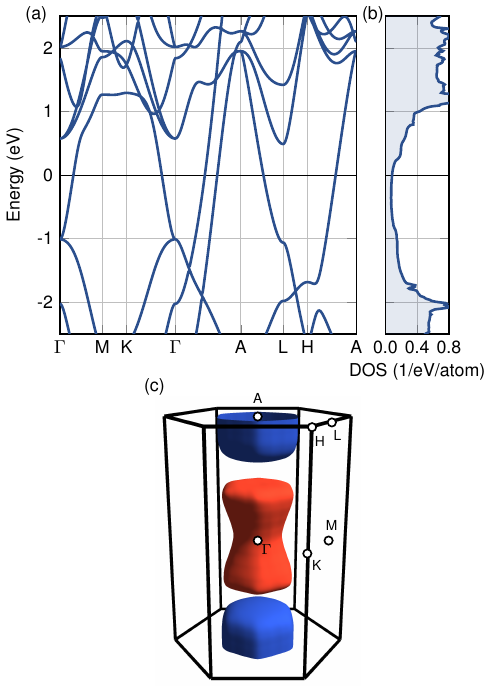}};
\end{tikzpicture}
\caption{(a) Band structure of $\ve$-TaN, showing highly dispersive bands crossing the Fermi level. (b) The corresponding electronic density of states $N_\text{F,at}$. (c) Fermi surface~\cite{Kawamura2019} of $\ve$-TaN, with the electron and hole pockets shown in red and blue, respectively.}
\label{f:bandfs}
\end{figure}

In Figs.~\ref{f:therm}(b) and (c), we separately compare the calculated electronic and lattice thermal conductivities with experiment. To obtain the electronic contribution, we measure the electrical conductivity $\sigma$ and estimate $\kappa_\text{el} = \sigma T L_0$, where $L_0$ is the Sommerfeld value of the Lorenz number~\cite{Kittel1976}. We then extract $\kappa_\text{ph} = \kappa \!-\! \kappa_\text{el}$. This method effectively separates the electronic and lattice contributions to the thermal conductivity in small-grained samples for which elastic scattering dominates. Appendix~\ref{app:convtests} discusses the theoretical calculation of the Lorenz ratio for bulk $\varepsilon$-TaN and $\varepsilon$-TaN samples with grain sizes of 0.1~\mum\ and 0.5~\mum. These calculations show that the Lorenz number is close to $L_0$ as a result of grain-boundary scattering.

The measurements (symbols) support the conclusion that $\kappa_{\rm el}$ and $\kappa_{\rm ph}$ both contribute substantially to the total thermal conductivity. Overall, these results suggest that reducing extrinsic scattering, for example with an increased grain size or in single crystals, could enable $\ve$-TaN to approach the thermal conductivity predicted by our calculations. 

\begin{figure}[t]
\centering
\begin{tikzpicture}[scale=1.0,every node/.style={scale=1.0}]
  \def\offset{0}
  \node[inner sep=0pt] at (0,0){\includegraphics[width=0.43\textwidth]{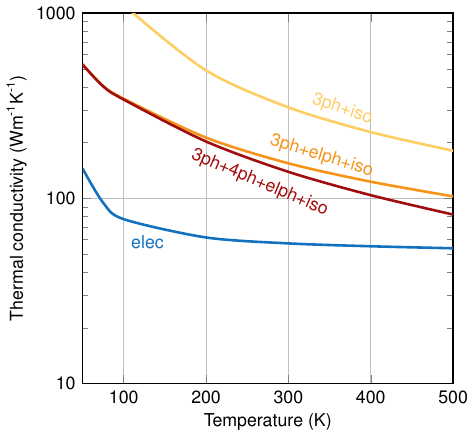}};
\end{tikzpicture}
\caption{Calculated isotropic average of the thermal conductivity of $\ve$-TaN as a function of temperature. The blue line is the electronic contribution. The yellow line is the calculation including only three-phonon scattering and isotope scattering (the latter reduces the lattice thermal conductivity by only 0.4\% of the total at most). The orange line includes also electron-phonon scattering, and the red line includes four-phonon scattering.}
\label{f:cont}
\end{figure}

\begin{figure}[t]
\centering
\begin{tikzpicture}[scale=1,every node/.style={scale=1.0}]
  \def\offset{0}
  \node[inner sep=0pt] at (0,0){\includegraphics[width=0.47\textwidth]{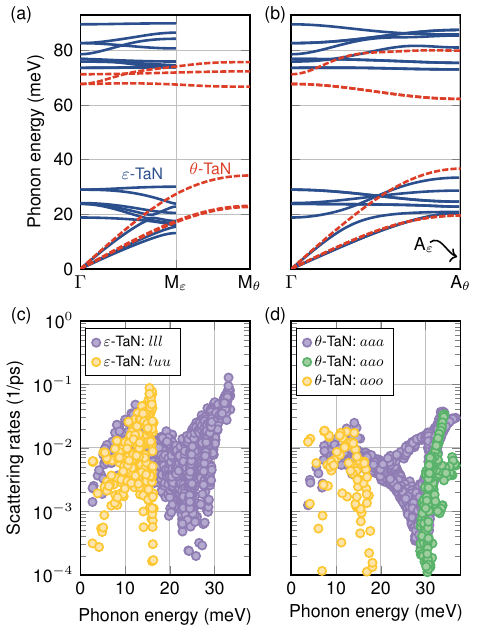}};
\end{tikzpicture}
\caption{(a) and (b) Phonon dispersion relations of $\ve$-TaN along the $\Gamma$-M and $\Gamma$-A high-symmetry lines (solid blue lines), compared to the corresponding dispersions of $\theta$-TaN (red dashed lines). The subscripts $\ve$ and $\theta$ in the labels of high-symmetry points serve as a reminder that the $\ve$-phase corresponds to a $\sqrt{3}\times\sqrt{3}$ supercell of the $\theta$-phase in the $ab$ plane. (c) and (d) Three-phonon scattering rates in $\ve$-TaN and $\theta$-TaN, decomposed by scattering process.}
\label{f:disc}
\end{figure}

We now analyze the origin of the electronic and lattice thermal conductivities in $\ve$-TaN. Figures~\ref{f:bandfs}(a) and (b) show the electronic band structure and density of states of $\ve$-TaN, respectively. The bands crossing the Fermi level are highly dispersive, with an average Fermi velocity of $0.7\cdot 10^6$\,m/s, more than twice that of the high-thermal-conductivity polymorph $\theta$-TaN (Tab.~\ref{t:fv}). This large Fermi velocity contributes to the sizable $\kappa_\text{el}$ shown as the blue line in Figure~\ref{f:cont}. Furthermore, the Fermi surface shown in Fig.~\ref{f:bandfs}(c) consists of a three-dimensional electron pocket centered at $\Gamma$ (red) and a three-dimensional hole pocket centered at $A$ (blue). This relatively simple topology suggests limited nesting, consistent with weak or moderate electron-phonon scattering.

Turning to the lattice thermal conductivity, The decomposition of phonon scattering rates is provided in Tab.~\ref{t:scattcomp}. Although an iterative solution of the BTE is required to accurately calculate the lattice thermal conductivity, the relaxation time approximation (RTA) remains useful for interpreting the roles of the various scattering mechanisms. We propose a method to average each scattering component using the phonon BTE kernel, as described in Appendix~\ref{app:avgscatt}. Figure~\ref{f:cont} shows the decomposition of the isotropically averaged $\kappa_\text{ph}$ into scattering channel. The main scattering mechanisms are three-phonon scattering and electron-phonon scattering. In particular, electron-phonon scattering reduces the lattice thermal conductivity by 50\% at room temperature, indicating that this mechanism is significant but still allows sizable heat transport by the lattice. This finding is consistent with the analysis of Eq.~\eqref{Eq:klat2} and with the small density of states at the Fermi level, 0.085/eV/atom, shown in Fig.~\ref{f:bandfs}(b). 

\begin{table*}
\caption{\label{t:scattcomp} Calculated lattice thermal conductivity and averaged phonon scattering rates of the $\varepsilon$, $\theta$, and $\delta$ phases of TaN, respectively, decomposed by scattering channel. In all cases we report values. As already reported in Ref.~\citenum{Kundu2021}, in the case of $\theta$-TaN, phonon scattering is dominated by three-phonon and four-phonon scattering, while the contribution of electron-phonon scattering to $\kappa_\text{ph}$ is negligible.
In the case of $\varepsilon$-TaN, the electron-phonon scattering rate is smaller than three-phonon scattering rates, but is no longer negligible.
For $\delta$-TaN, electron-phonon scattering is the dominant scattering channel, and leads to a very low lattice thermal conductivity. The calculation undertainty is estimated to be $\pm 5$\wmk.}
\begin{ruledtabular}
{\renewcommand{\arraystretch}{1.2}
\begin{tabular}{lccccc}
\textrm{\ }&\\[-13pt]
phase & \textrm{$\kappa_{\text{ph,RTA}}$}&
\textrm{$\big< 1/ \tau^{\text{(3ph)}}_\text{ph}\big> $}&
\textrm{$\big< 1/ \tau^{\text{(4ph)}}_\text{ph}\big> $}&
\textrm{$\big< 1/ \tau^{\text{(elph)}}_\text{ph}\big>$}&
\textrm{$\big< 1/ \tau^{\text{(iso)}}_\text{ph}\big> $}\\[3pt]
 & (Wm$^{-1}$K$^{-1})$&
(ps$^{-1})$&
(ps$^{-1})$&
(ps$^{-1})$&
(ps$^{-1})$\\[4pt]
\colrule\\[-11pt]
$\varepsilon$ & 116 & 1.1$\times$10$^{\text{-2}}$ & 8.5$\times$10$^{\text{-4}}$ & 3.8$\times$10$^{\text{-3}}$ & 2.2$\times$10$^{\text{-4}}$\\
$\theta$ & 463 & 5.4$\times$10$^{\text{-3}}$ & 8.9$\times$10$^{\text{-4}}$ & 7.3$\times$10$^{\text{-4}}$ & 2.1$\times$10$^{\text{-4}}$\\
$\delta$ & 11 & 4.4$\times$10$^{\text{-2}}$ & 3.5$\times$10$^{\text{-2}}$ & 1.3$\times$10$^{\text{-1}}$ & 1.4$\times$10$^{\text{-3}}$\\
\end{tabular}
}
\end{ruledtabular}
\end{table*}

Despite the reduction of $\kappa_\text{ph}$ by both three-phonon and electron-phonon scattering, the lattice thermal conductivity remains unusually high for a metal. To clarify this observation, we inspect the phonon dispersion relations in Figs.~\ref{f:disc}(a) and (b) and compare to those of $\theta$-TaN. The unit cell of $\ve$-TaN contains six atoms, leading to 18 phonon branches, and can be viewed as a distorted $\sqrt{3}\times\sqrt{3}\times1$ supercell of $\theta$-TaN; see Figs.~~\ref{f:struc}(a) and (b). Accordingly, the phonon dispersions of $\ve$-TaN can be interpreted approximately in terms of zone folding from the $\theta$ phase. 

This relationship has two important consequences. First, $\ve$-TaN retains a high speed of sound, comparable to that of $\theta$-TaN ($6.5\cdot 10^3$\,m/s vs.\ $7.0\cdot 10^3$\,m/s) favoring high $\kappa_\text{ph}$, as shown by Eq.~\eqref{Eq:klat2}. Second, $\ve$-TaN exhibits a phonon gap that is even wider than in $\theta$-TaN (39.6\,meV vs.\ 25.9\,meV) restricting the phase space for three-phonon scattering. In $\ve$-TaN, this gap does not separate acoustic and optical modes in the usual sense. Rather, it separates a lower group of nine branches ($l$), arising from the folding of the acoustic modes of $\theta$-TaN, from an upper group of nine branches ($u$), arising from the folding of the optical modes. Owing to this phonon gap, $llu$ processes are energetically forbidden: the maximum frequency in the $l$ manifold is smaller than the minimum frequency in the $u$ manifold. As a result, the three-phonon scattering that degrades phonon transport is dominated by $lll$ and $luu$ processes, as shown in Fig.~\ref{f:disc}(c). This behavior is analogous to the suppression of $aao$ processes in $\theta$-TaN, shown in Fig.~\ref{f:disc}(d).

We performed the same analysis for $\delta$-TaN as for $\varepsilon$-TaN. As shown in Fig.~\ref{f:map} and Table~\ref{t:scattcomp}, $\delta$-TaN exhibits lower electronic and lattice thermal conductivities, comparable to those of typical metals, due of strong electron-phonon and phonon-phonon scattering. The temperature-dependent electronic and lattice thermal conductivities, phonon dispersion including anharmonic effects, electronic band structure, density of states, and Fermi surface of $\delta$-TaN are discussed in Appendix~\ref{app:delta}.


We now return to the scaling analysis of Eqs.~\eqref{Eq:kel2} and \eqref{Eq:klat2}. For use in Eqs.~\eqref{Eq:kel2} and \eqref{Eq:klat2}, we evaluated the characteristic electron-phonon matrix element $\tilde{g}$ defined in Eqs.~\eqref{eq.gamma-allen3} and \eqref{eq.gamma-allen4} for the three TaN phases using \textit{ab initio} calculations. The value of $\tilde{g}$ was obtained using the same computational parameters as those used to calculate the electron-phonon scattering rates entering the lattice thermal conductivity. The calculated values of $\tilde{g}$ for $\varepsilon$-, $\theta$-, and $\delta$-TaN are 58, 66, and 125~meV, respectively. These results indicate that the electron-phonon interaction is rather weak in $\varepsilon$-TaN; in fact, our calculation of the dimensionless electron-phonon coupling strength, which also includes information about the scattering phase space, yields the very small value $\lambda=0.07$.

Finally, using the values of $N_{\rm F,at}$, $v_\text{s}$, $v_\text{F}$, and $\tilde g$ reported above, and taking the maximum energies of the relevant acoustic or lower phonon manifolds to be $\hbar\omega \sim 30$~meV, $\hbar\omega \sim 35$~meV, and $\hbar\omega \sim 25$~meV for $\varepsilon$-, $\theta$-, and $\delta$-TaN, respectively, Eq.~\eqref{Eq:kel2} yields estimates of 55\wmk, 31\wmk, and 69\wmk, respectively. The corresponding isotropically averaged room-temperature electronic thermal conductivities obtained from full \textit{ab initio} calculations are 57\wmk, 40\wmk, and 64\wmk, respectively. These results show that the scaling expression provides reasonable estimates of the electronic thermal conductivity. We note that the numerical estimates include a prefactor of $\pi/36$, which was omitted from Eq.~\eqref{Eq:kel2} for simplicity during the derivation.

On the other hand, the upper bounds on the lattice thermal conductivity estimated from Eq.~\eqref{Eq:klat2} are 754\wmk, 2068\wmk, and 121\wmk\ for $\varepsilon$-, $\theta$-, and $\delta$-TaN, respectively. These values are in good agreement with the trend of the RTA lattice thermal conductivities reported in Tab.~\ref{t:scattcomp}. (We note that these numerical estimates include a prefactor of $1/\pi$, which was omitted from Eq.~\eqref{Eq:klat2} for simplicity during the derivation). However, this scaling estimate has limitations: it can deviate from the full \textit{ab initio} results at high temperatures which lead to strong phonon-phonon interaction or in materials with strong intrinsic phonon-phonon scattering. Although the proposed scaling relation cannot replace full \textit{ab initio} calculations, the comparison among the three TaN phases confirms that it provides a useful guide for identifying materials with high thermal conductivity.

\section{Conclusion}\label{sec:conclusion}
In summary, we have identified $\ve$-TaN as a rare material in which electronic and lattice heat transport are both unusually efficient. Our combined scaling analysis and \textit{ab initio} calculations show that this behavior originates from the coexistence of a large Fermi velocity and a low density of states on the electronic side, together with a large speed of sound and a wide phonon gap on the lattice side. These predictions are supported by our experimental synthesis and characterization of $\ve$-TaN, which confirm sizable contributions from both electrons and phonons to the thermal conductivity of this compound.

More broadly, our work points to a design strategy for high-$\kappa$ metallic materials based on the joint optimization of $v_\text{F}^2$ and $N_{\rm F,at}^{-1}/v_\text{F}$. It also highlights a relatively unexplored regime of heat transport in which electrons and phonons contribute in comparable measure.\\

\begin{acknowledgments}
We are grateful to Keivan Esfarjani for fruitful discussions. This work was supported by the U.S. Department of Energy's Office of Energy Efficiency and Renewable Energy (EERE) under the Industrial Efficiency and Decarbonization Office Award Number DE-EE0011229 (calculations and experiments); and by the Computational Materials Science program of the U.S. Department of Energy, Office of Science, Basic Energy Sciences, through award no. DE-SC0020129 (development of electronic thermal conductivity module in \textsc{EPW}). Computational resources were provided by the National Energy Research Scientific Computing Center (a DOE Office of Science User Facility supported under Contract No.~DE-AC02-05CH11231), the Argonne Leadership Computing Facility (a DOE Office of Science User Facility supported under Contract DE-AC02-06CH11357), and the Texas Advanced Computing Center (TACC) at The University of Texas at Austin.
\end{acknowledgments}

\appendix

\section{Derivation of the scaling laws}\label{app:derive}
Here we outline the derivation of the scaling laws for the electronic thermal conductivity and the lattice thermal conductivity given by Eqs.~(\ref{Eq:kel2}) and (\ref{Eq:klat2}). In both cases, the key is to derive simple expressions for the electron relaxation times and the phonon relaxation times resulting from electron-phonon interactions, namely Eqs.~(\ref{Eq:tauel}) and (\ref{Eq:tauph}), respectively. We start from the electronic relaxation times.

\subsection{Electron relaxation time from electron-phonon coupling}
The relaxation time $\tau_{n\bf k}$ of an electron in the Kohn-Sham state with band index $n$ and wavevector $\bf k$ is obtained from the imaginary part of the Fan-Midgal self-energy and is given by~\cite{Giustino2017}:
  \begin{eqnarray}
  \frac{1}{\tau_{n\bk}} &=& \frac{2\pi}{\hbar}
  \sum_{m\nu} \!\int\!\! \frac{d\bq}{\Omega_{\rm BZ}} |g_{nm\nu}(\bk,\bq)|^2 \nonumber \\
  &\times&\left[ (1-f_{m\bk+\bq}+n_{\bq\nu})\d(\ve_{n\bk} -\hbar\w_{\bq\nu}-\!\ve_{m\bk+\bq})\right.  \nonumber \\
  &+&\left.(f_{m\bk+\bq} +n_{\bq\nu})\,\,\d(\ve_{n\bk}+\hbar\w_{\bq\nu} -\!\ve_{m\bk+\bq})\right]\!.\,\,\, 
  \label{eq.fermirule-gamma}
  \end{eqnarray}
Here, $\hbar$ is the Planck constant, and $\Omega_{\rm BZ}$ the volume of the Brillouin zone; $\Omega_{\rm BZ} = (2\pi)^3/\Omega$, where $\Omega$ is the volume of the primitive unit cell.
$g_{mn\nu}(\bk,\bq)$ is the electron-phonon matrix element for the transition between the Kohn-Sham states $n\bk$ and $m\bk+\bq$ via the phonon $\bq\nu$. $\ve_{n\bk}$ is the electron band energy and $\w_{\bq\nu}$ is the phonon frequency. $f_{n\bk}$ and $n_{\bq\nu}$ denote Fermi-Dirac and Bose-Einstein thermal occupations, respectively. At high temperature ($k_{\rm B}T \gg \hbar\w_{\bq\nu}$, with $k_{\rm B}$ being the Boltzmann constant and $T$ the absolute temperature), we can approximate the occupation factor $(1-f_{m\bk+\bq}+n_{\bq\nu})$ with $n_{\bq\nu}$, and similarly for $(f_{m\bk+\bq} +n_{\bq\nu})$. Furthermore, in the same limit, the Bose-Einstein distribution can be replaced by $k_{\rm B}T/\hbar\w_{\bq\nu}$. Under these approximations, Eq.~\eqref{eq.fermirule-gamma} reduces to:
  \begin{eqnarray}
  \frac{1}{\tau_{n\bk}} &\simeq& \frac{2\pi}{\hbar}
  \sum_{m\nu} \!\int\!\! \frac{d\bq}{\Omega_{\rm BZ}} |g_{nm\nu}(\bk,\bq)|^2 \frac{k_{\rm B}T}{\hbar\w_{\bq\nu}}\nonumber \\
  &\times&\left[ \d(\ve_{n\bk} -\hbar\w_{\bq\nu}-\!\ve_{m\bk+\bq})\right. \nonumber\\
  &+&\left.\d(\ve_{n\bk}+\hbar\w_{\bq\nu} -\!\ve_{m\bk+\bq})\right]\!.\,\,\, 
  \label{eq.fermirule-gamma2}
  \end{eqnarray}
Since in Eq.~(\ref{Eq:kel}) we need the average electron relaxation time in the vicinity of the Fermi surface, we can neglect the phonon energy in the Dirac delta functions of the above expression, obtaining:
  \begin{eqnarray}
  \frac{1}{\tau_{n\bk}} &\simeq& \frac{4\pi k_{\rm B}T}{\hbar}
  \sum_{m\nu} \!\int\!\! \frac{d\bq}{\Omega_{\rm BZ}}  \frac{|g_{nm\nu}(\bk,\bq)|^2}{\hbar\w_{\bq\nu}}\nonumber\\
  &\times&\d(\ve_{n\bk} -\ve_{m\bk+\bq})~. 
  \label{eq.fermirule-gamma3}
  \end{eqnarray}
Equation~(\ref{Eq:tauel}) follows from this expression by factoring out the density of electronic states at the Fermi energy per unit volume, $N_{\rm F} = \Omega_{\rm BZ}^{-1}\sum_n\int \!d\bk \,\d(\ve_{\rm F} -\ve_{n\bk})/\Omega$:
\begin{widetext}
  \begin{eqnarray}
  \frac{1}{\tau_{n\bk}} &\simeq& 4\pi \frac{\Omega N_{\rm F}}{\hbar} k_{\rm B}T
  \,\frac{ \!\Omega_{\rm BZ}^{-1}\sum_{m}\int\!\! d\bq\,\,  \left[\sum_\nu |g_{nm\nu}(\bk,\bq)|^2/\hbar\w_{\bq\nu}\right]\d(\ve_{n\bk} -\ve_{m\bk+\bq})}{\Omega_{\rm BZ}^{-1}\sum_m\int \!\!d\bq\,\, \d(\ve_{\rm F} -\ve_{m\bk+\bq})}~,
  \label{eq.fermirule-gamma4}
  \end{eqnarray}
\end{widetext}
and by identifying the average coupling strength $g^2/\hbar\w$ with the ratio between the two integrals on the right-hand side.

We note that, In Eq.~(\ref{Eq:tauel}), the product $\Omega\, g^2$ is an intensive quantity owing to the normalization of vibrational eigenmodes. 

\subsection{Phonon relaxation time from electron-phonon coupling}
The relaxation time $\tau_{\bq\nu}$ of a phonon with wavevector $\bq$ and branch index $\nu$ is obtained from the imaginary part of the phonon self-energy and is given by~\cite{Giustino2017}:
  \begin{eqnarray}
  \frac{1}{\tau_{\bq\nu}} &=& 
  \frac{2\pi}{\hbar} 2 \sum_{mn} \!\int\! \!\frac{d\bk}{\Omega_{\rm BZ}} |g_{mn\nu}(\bk,\bq)|^2\nonumber\\
  &\times&(f_{n\bk}-f_{m\bk+\bq})\d(\ve_{m\bk+\bq}-\ve_{n\bk}-\hbar\w_{\bq\nu})\,,
      \label{eq.gamma-allen}
  \end{eqnarray}
where the meaning of the symbols is the same as in the previous section and the factor of 2 accounts for the spin degeneracy. In this case, the expression can be simplified by noting that the occupation factor $(f_{n\bk}-f_{m\bk+\bq})$ is peaked near the Fermi energy, therefore we can write~\cite{Allen1972}:
  \begin{eqnarray}
  \frac{1}{\tau_{\bq\nu}} &\simeq& 
  \frac{2\pi}{\hbar} 2 \hbar\w_{\bq\nu}\sum_{mn} \!\int\! \!\frac{d\bk}{\Omega_{\rm BZ}} |g_{mn\nu}(\bk,\bq)|^2
  \nonumber\\
  &\times& \left(-\frac{\partial f_{n\bk}}{\partial\ve_{n\bk}}\right) \d(\ve_{m\bk+\bq}-\ve_{n\bk}-\hbar\w_{\bq\nu})~.
      \label{eq.gamma-allen2}
  \end{eqnarray}
The derivative and the Dirac delta function on the right can further be approximated by two Dirac deltas centered at the Fermi energy:
  \begin{eqnarray}
  \frac{1}{\tau_{\bq\nu}} &\simeq& 
  4\pi\w_{\bq\nu}\sum_{mn} \!\int\! \!\frac{d\bk}{\Omega_{\rm BZ}} |g_{mn\nu}(\bk,\bq)|^2\nonumber\\
  &\times& \d(\ve_{n\bk}-\ve_{\rm F})\d(\ve_{m\bk+\bq}-\ve_{\rm F})~.
      \label{eq.gamma-allen3}
  \end{eqnarray}
In analogy with Eq.~\eqref{eq.fermirule-gamma4}, an average electron-phonon coupling can be defined by factoring out the Fermi surface nesting function $\chi_\bq = 2 \Omega_{\rm BZ}^{-1}\sum_{mn} \!\int\! \!d\bk\,   \d(\ve_{n\bk}-\ve_{\rm F})\d(\ve_{m\bk+\bq}-\ve_{\rm F})$:
\begin{widetext}
  \begin{equation}
  \frac{1}{\tau_{\bq\nu}} \simeq 
  2\pi\w_{\bq\nu} \frac{\Omega_{\rm BZ}^{-1}\sum_{mn} \!\int\! \!d\bk\, |g_{mn\nu}(\bk,\bq)|^2
  \d(\ve_{n\bk}-\ve_{\rm F})\d(\ve_{m\bk+\bq}-\ve_{\rm F})}{\Omega_{\rm BZ}^{-1}\sum_{mn} \!\int\! \!d\bk\, 
  \d(\ve_{n\bk}-\ve_{\rm F})\d(\ve_{m\bk+\bq}-\ve_{\rm F})}\, \chi_\bq~,
      \label{eq.gamma-allen3}
  \end{equation}
\end{widetext}
and identifying the weighted average with the characteristic electron-phonon matrix element $\tilde{g}$. We use the tilde symbold to distinguish this quantity from the electronic $g$ defined by Eq.~\eqref{eq.fermirule-gamma4}. Under these approximations, we find the compact expression:
  \begin{equation}
  \frac{1}{\tau_{\bq\nu}} \simeq 
  2\pi\w_{\bq\nu}\, {\tilde g}^2\, \chi_\bq~.
      \label{eq.gamma-allen4}
  \end{equation}
In general, the nesting function must be evaluated numerically. However, to obtain a simple scaling relation we can use the textbook result for the electron gas, which is:   \begin{equation}
  \chi_\bq = \frac{N_{\rm F} \Omega}{2\hbar v_{\rm F}|\bq|} \quad \mbox{for} \quad |\bq| < 2k_{\rm F},
      \label{eq.nesting}
  \end{equation}
and vanishes otherwise. Here, $v_{\rm F}$ and $k_{\rm F}$ are the Fermi velocity and Fermi wavevector, respectively. Now we focus on the acoustic modes since they dominate heat transport; in this case, we can approximate the phonon dispersion relations via $\omega_{\bq\nu} = v_\text{s}|\bq|$, with $v_{\rm s}$ being the speed of sound. Using this substitution in Eq.~\eqref{eq.gamma-allen4}, and combining with Eq.~\eqref{eq.nesting}, we obtain Eq.~(\ref{Eq:tauph}).

\section{Convergence tests}\label{app:convtests}
\subsection{Wannier interpolation}\label{app:convtests-wannier}
\begin{figure}[t]
\centering
\begin{tikzpicture}[scale=1,every node/.style={scale=1.0}]
  \def\offset{0}
  \node[inner sep=0pt] at (0,0){\includegraphics[width=0.47\textwidth]{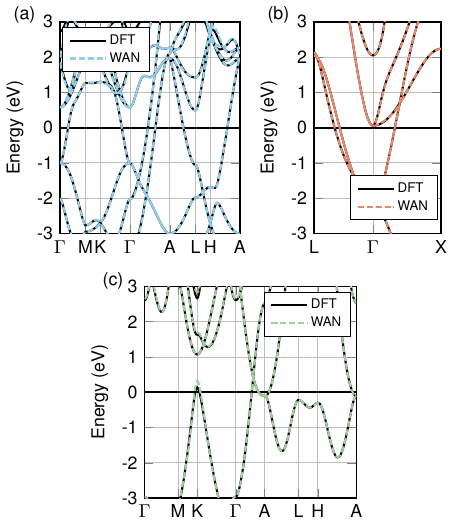}};
\end{tikzpicture}
\caption{Wannier-interpolated band structures of TaN polymorphs. (a) DFT bands of $\ve$-TaN (black solid lines) and Wannier-interpolated bands (blue dashed lines). (b) DFT bands of $\delta$-TaN (black solid lines) and Wannier-interpolated bands (orange dashed lines). (c) DFT bands of $\theta$-TaN (black solid lines) and Wannier-interpolated bands (green dashed lines).}
\label{f:bands}
\end{figure}
For the Wannier-Fourier interpolation in the \textsc{EPW} code, we employ 9 Wannier functions for $\varepsilon$-TaN, starting from $d_{z^{2}}$, $d_{xz}$, and $d_{yz}$ orbitals on one Ta1 atom, $d_{z^{2}}$, $d_{x^{2}-y^{2}}$, and $d_{xy}$ orbitals on one Ta2 atom, and $p$ orbitals on one N atom. The labels for the Ta atoms are indicated in Fig.~\ref{f:struc}. In the case of $\delta$-TaN, we employ 9 Wannier functions, starting from $d$ orbitals on Ta atom and $p$, and $s$ orbitals on N atom. In the case of $\delta$-TaN, we employ 9 Wannier functions, starting from $d$ orbitals on Ta atom and $p$, and $s$ orbitals on N atom. Figure~\ref{f:bands} shows a comparison between our calculated DFT band structures and the bands obtained from Wannier interpolation using these choices.

\begin{figure}[t]
\centering
\begin{tikzpicture}[scale=1,every node/.style={scale=1.0}]
  \def\offset{0}
  \node[inner sep=0pt] at (0,0){\includegraphics[width=0.42\textwidth]{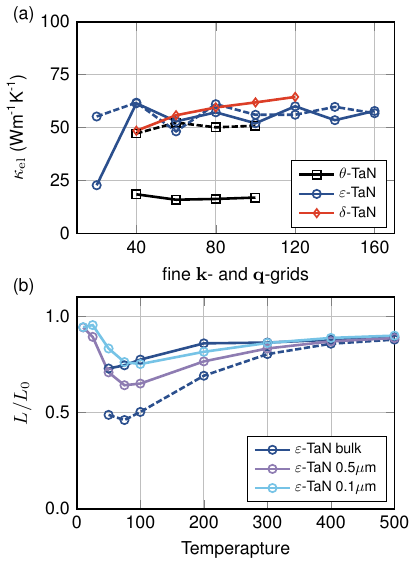}};
\end{tikzpicture}
\caption{(a) Convergence tests for the electronic thermal conductivity with respect to the Brillouin-zone sampling used in the BTE calculations, for all three polymorphs of TaN. Solid and dashed lines represent $a$-axis and $c$-axis values at room temperature, respectively. In case of $\varepsilon$-TaN, the corresponding samplings along the $a$-axis are 3/5 of these values. (b) Calculated Lorenz ratio of bulk, 0.5~\mum, and 0.1~\mum\ $\varepsilon$-TaN as a function of temperature. For cases including grain-boundary scattering, the Lorenz ratios are isotropically averaged. In the cases of $\ve$-TaN and $\theta$-TaN, the horizontal axis indicates the number of grid points along the $c$ axis, and dashed and solid lines represent the $a$- and $c$-axis values, respectively.}
\label{f:elconv}
\end{figure}

\begin{figure*}[t]
\centering
\begin{tikzpicture}[scale=1,every node/.style={scale=1.0}]
  \def\offset{0}
  \node[inner sep=0pt] at (0,0){\includegraphics[width=0.96\textwidth]{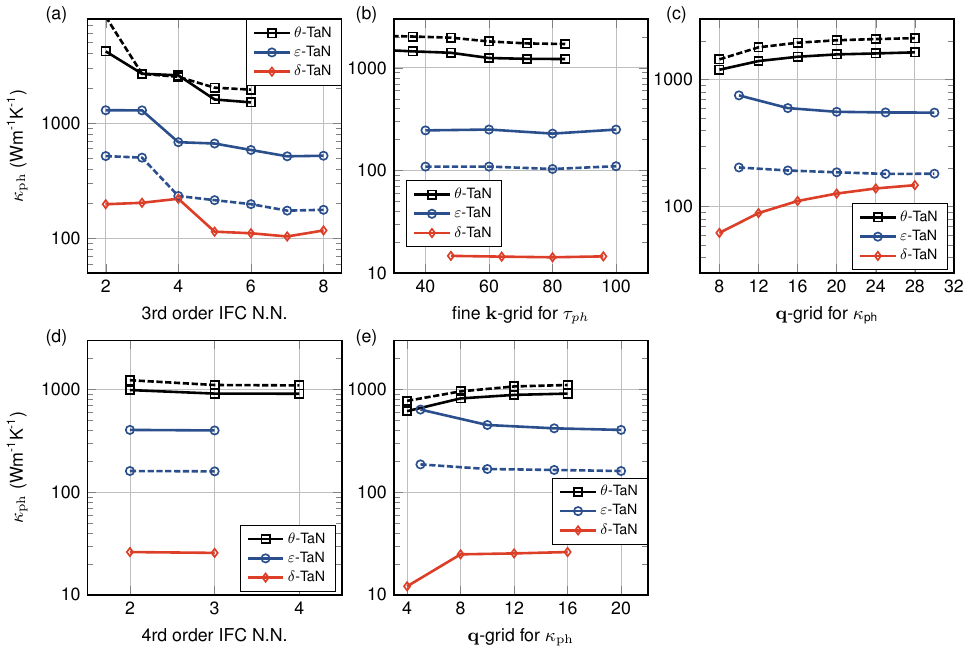}};
\end{tikzpicture}
\caption{Convergence tests for the lattice thermal conductivity, for all three polymorphs of TaN. (a) Convergence of the room-temperature thermal conductivity with respect to the number of nearest-neighbor (N.N.) shells included in the third-order force constants. (b) Convergence with respect to the Brillouin-zone sampling used to compute $\tau_\text{ph}$. (c) Three-phonon scattering convergence with respect to the Brillouin-zone sampling used to compute $\kappa_\text{ph}$. (d) Convergence of the room-temperature thermal conductivity with respect to the number of N.N. shells included in the fourth-order force constants. (e) Four-phonon scattering convergence with respect to the Brillouin-zone sampling used to compute $\kappa_\text{ph}$. Solid and dashed curves denote transport along the
$a$ and $c$ axes, respectively. Since $\ve$-TaN and $\theta$-TaN are anisotropic, in both cases we show the $c$-axis values as solid lines, and the $a$-axis values as dashed lines.
In these anisotropic cases, the horizontal axis indicates the grids along the $c$ axis (the corresponding samplings along the $a$-axis are 3/5 of these values).
Panels (a) and (c) include only three-phonon and isotope scattering; panel (b) also
includes electron-phonon scattering. Panels (d) and (e) include three-phonon, four-phonon, and isotope scattering.}
\label{f:latconv}
\end{figure*}

\subsection{Electronic thermal conductivity}\label{app:convtests-eltherm}
Figure~\ref{f:elconv}(a) reports the convergence of the electronic thermal conductivity $\kappa_{\mathrm{el}}$ in $\varepsilon$-TaN, $\delta$-TaN, and $\theta$-TaN as a function of the electronic Brillouin-zone sampling used to evaluate the transport integrals within the electronic BTE. The calculations show that $96\times96\times160$, $120^3$, and $100^3$ $\mathbf{k}$-point grids are sufficient to converge $\kappa_{\mathrm{el}}$ within 5\wmk.

Figure~\ref{f:elconv}(b) shows the calculated Lorenz ratios of $\varepsilon$-TaN for different grain sizes. Bulk $\varepsilon$-TaN exhibits a relatively lower Lorenz ratio than calculations that include grain-boundary scattering. This suggests that, in the bulk material, energy and momentum relaxation affect heat and charge transport differently. However, the Lorenz ratio increases for a grain size of 0.5~\mum\ and rises further for a grain size of 0.1~\mum. In particular, the increase in the Lorenz ratio at low temperatures (25-50~K) indicates that elastic grain-boundary scattering is the dominant scattering mechanism.

\begin{figure*}[t]
\centering
\begin{tikzpicture}[scale=1,every node/.style={scale=1.0}]
  \def\offset{0}
  \node[inner sep=0pt] at (0,0){\includegraphics[width=0.7\textwidth]{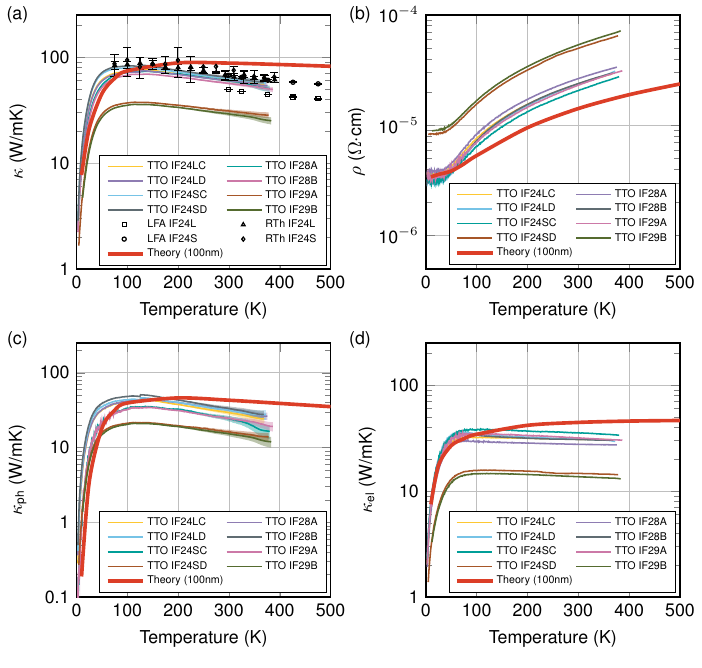}};
\end{tikzpicture}
\caption{Comparison between measured and calculated electrical and thermal conductivity of $\ve$-TaN. (a) Total thermal conductivity. (b) Electrical resistivity. (c) Lattice thermal conductivity. (d) Electronic thermal conductivity. Measurements are reported for all samples indicated in Tab.~\ref{t:samplelist}, as indicated in the legend. For clarity of visualization, we only report the theoretical values corresponding to the isotropic average and including grain-boundary scattering calculated with a 100~nm grain size.}
\label{f:fulltherm}
\end{figure*}

\begin{table}[b]
\caption{\label{t:rttherm} Measured average lattice ($\kappa_\text{ph}$), electronic ($\kappa_\text{el}$), and total thermal ($\kappa$) conductivity at room-temperature for three different $\varepsilon$-TaN composites.\vspace{3pt}}
\begin{ruledtabular}
{\renewcommand{\arraystretch}{1.2}
\begin{tabular}{lcccc}
\textrm{Sample}&
\textrm{Method}&
\textrm{$\kappa_\text{ph}$}&
\textrm{$\kappa_\text{el}$}&
\textrm{$\kappa$}\\ 
\textrm{\ }&
\textrm{\ }&
(Wm$^{-1}$K$^{-1}$)&
(Wm$^{-1}$K$^{-1}$)&
(Wm$^{-1}$K$^{-1}$)\\[4pt] 
\colrule\\[-11pt]
IF24S & TTO & 33 $\pm$ 3 & 31 $\pm$ 1 & 63 $\pm$ 3 \\
IF24S & RTh & - & - & 67 $\pm$ 7 \\
IF24S & LFA & - & - & 67 $\pm$ 2 \\
IF24S & psTTR & - & - & 130 $\pm$ 9 \\
IF24S & psTTR & - & - & 15 $\pm$ 2 \\
IF24L & TTO & 31 $\pm$ 2 & 29 $\pm$ 2 & 60 $\pm$ 2 \\
IF24L & RTh & - & - & 68 $\pm$ 6 \\
IF24L & LFA & - & - & 49 $\pm$ 1 \\
IF28 & TTO & 24 $\pm$ 2 & 34 $\pm$ 2 & 58 $\pm$ 2 \\
IF29 & TTO & 16 $\pm$ 1 & 14 $\pm$ 1 & 30 $\pm$ 1 \\
\end{tabular}
}
\end{ruledtabular}
\end{table}

\begin{figure*}[t]
\centering
\begin{tikzpicture}[scale=1,every node/.style={scale=1.0}]
  \def\offset{0}
  \node[inner sep=0pt] at (0,0){\includegraphics[width=0.92\textwidth]{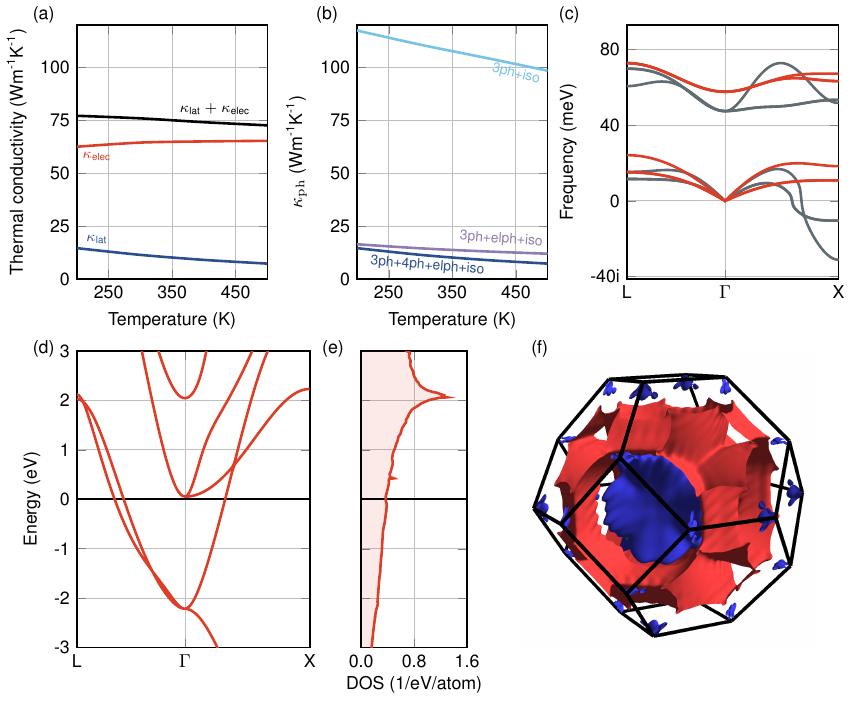}};
\end{tikzpicture}
\caption{Theoretical analysis of $\delta$-TaN. (a) Calculated lattice, electronic, and total thermal conductivities. (b) Calculated lattice thermal conductivity. The skyblue line is the calculation including only three-phonon scattering and isotope scattering. The purple line includes also electron-phonon scattering, and the blue line includes four-phonon scattering. (c) Phonon dispersions calculated within the harmonic approximation (gray) and including anharmonic corrections (red). (d) Electronic band structure and (e) electronic density of states (DOS). (f) Fermi surface of $\delta$-TaN~\cite{Kawamura2019}.}
\label{f:delta}
\end{figure*}

\subsection{Lattice thermal conductivity}\label{app:convtests-lattherm}

In Fig.~\ref{f:latconv} we analyze the convergence of the lattice thermal conductivity $\kappa_{\text{ph}}$ with respect to the number of neighbors in calculations of third-order force constants, number of $\bf k$-points in calculations of phonon relaxation times, and number of $\bf q$-points.
In particular, Fig.~\ref{f:latconv}(a) reports the convergence of the lattice thermal conductivity in $\varepsilon$-TaN, $\delta$-TaN, and $\theta$-TaN as a function of the number of nearest neighbor shells included in calculations of third-order interatomic force constants. The calculations show that 8 nearest-neighbor shells are sufficient to converge $\kappa_{\text{ph}}$ for $\ve$-TaN, and 6 nearest-neighbor shells are sufficient to converge $\kappa_{\text{ph}}$ for both $\theta$-TaN and $\delta$-TaN.
Figure~\ref{f:latconv}(b) shows how the lattice thermal conductivity (including three-phonon, phonon-isotope, and electron-phonon scattering) depends on the sampling of the electronic Brillouin zone. Based on this test, we choose a sampling of $72\times72\times120$ points for $\varepsilon$-TaN, $96^3$ points for $\delta$-TaN, and $84^3$ points for $\theta$-TaN.
Using these converged values, in Fig.~\ref{f:latconv}(c) we investigate the sensitivity of $\kappa_{\text{ph}}$ to the $\bf q$-grid, by considering three-phonon and phonon-isotope scattering. 

Fig.~\ref{f:latconv}(d) shows the convergence of the lattice thermal conductivity in $\varepsilon$-TaN, $\delta$-TaN, and $\theta$-TaN as a function of the number of nearest neighbor shells included in calculations of fourth-order interatomic force constants. The calculations show that 3 nearest-neighbor shells are sufficient to converge $\kappa_{\text{ph}}$ for $\theta$-TaN, and 2 nearest-neighbor shells are sufficient to converge $\kappa_{\text{ph}}$ for both $\ve$-TaN and $\delta$-TaN.
With chosen interatomic force constants, Fig.~\ref{f:latconv}(e) we investigate the sensitivity of $\kappa_{\text{ph}}$ to the $\bf q$-grid, by considering three-phonon, four-phonon, and phonon-isotope scattering. Based on Figs.~\ref{f:latconv}(c) and (e), we see that calculations are converged for grids with $12\times12\times20$ and $16\times16\times16$ points in the case of the $\varepsilon$ and $\theta$ phase, respectively. In the cases of the $\delta$, we employ a $16^3$-grid for computational convenience.

We estimate the uncertainty of the calculated lattice thermal conductivities using the last two points in each plot. The maximum uncertainty is 5\wmk, therefore we consider this to be the nominal uncertainty in all our calculated data.

\section{Measured and calculated electrical and thermal conductivity of $\ve$-TaN}\label{app:therm}
We synthesized three samples, IF24, IF28, and IF29, and performed four types of measurements, TTO, LFA, RTh, and psTTR, on the IF24 sample. For IF28 and IF29, only TTO measurements were performed. All measured values are compared with theoretical calculations for 100~nm grains in Fig.~\ref{f:fulltherm}. Representative room-temperature thermal conductivity values are listed in Table~\ref{t:rttherm}, with the lattice and electronic contributions to the thermal conductivity specified separately for the TTO measurements.

\section{Average scattering rates for lattice thermal conductivity}\label{app:avgscatt}
To quantify the contribution of each scattering mechanism to the lattice thermal conductivity, we evaluate averaged scattering rates as follows.

Within the relaxation time approximation (RTA), the lattice thermal conductivity from the phonon BTE is:
\begin{equation}
\kappa_{\text{RTA}, \alpha \beta}= \sum_{\nu} \int \frac{d^{3}\mathbf{q}}{\Omega_{BZ}}  C_{\text{ph,}\mathbf{q}\nu}  v_{\mathbf{q}\nu \alpha} v_{\mathbf{q}\nu \beta} \tau_{\text{ph,}\mathbf{q}\nu} 
\label{eq.rta},
\end{equation}
where $C_{\text{ph,}\mathbf{q}\nu}$ is the contribution to lattice heat capacity per unit volume from the phonon with wavevector $\mathbf{q}$, branch $\nu$, and group velocity ${\bf v}_{\mathbf{q}\nu}$. $\tau_{\text{ph,}\mathbf{q}\nu}$ is the phonon relaxation time, as obtained via Matthiessen's rule :
\begin{equation}
\frac{1}{\tau_{\text{ph,}\mathbf{q}\nu}}= \frac{1}{\tau_{\mathbf{q}\nu}^{\text{(3ph)}}}+\frac{1}{\tau_{\mathbf{q}\nu}^{\text{(4ph)}}}+\frac{1}{\tau_{\mathbf{q}\nu}^{\text{(elph)}}}+\frac{1}{\tau_{\mathbf{q}\nu}^{\text{(iso)}}}
\label{eq.matt}.
\end{equation}
In order to account for the relative weight of different phonon modes to the thermal conductivity, we evaluate the average scattering rate for each mechanism ``X'' (where ``X'' is ``3ph'', ``4ph'', ``elph'', or ``iso'') as:
\begin{eqnarray}
\left\langle\frac{1}{\tau_{\text{ph}}^{\text{(X)}}}\right\rangle_{\!\!\alpha\beta}&=& \left[\kappa_{\text{RTA}}\right]_{\alpha\gamma}^{-1} \sum_{\nu} \int \frac{d^{3}\mathbf{q}}{\Omega_\text{BZ}}  C_{\text{ph,}\mathbf{q}\nu}  \nonumber\\
&\times&  v_{\mathbf{q}\nu \gamma} v_{\mathbf{q}\nu \beta} \tau_{\text{ph,}\mathbf{q}\nu}\times \frac{1}{\tau_{\mathbf{q}\nu}^{\text{(X)}}}
\label{eq.avgscatt}.
\end{eqnarray}
With this definition, the RTA lattice thermal conductivity can be expressed \textit{exactly} via a generalized, anisotropic Matthiessen's formula:
\begin{equation}
\kappa_{\text{RTA}, \alpha \beta}= \sum_{\nu} \int \frac{d^{3}\mathbf{q}}{\Omega_{BZ}}  C_{\text{ph,}\mathbf{q}\nu}  v_{\mathbf{q}\nu \alpha} v_{\mathbf{q}\nu \gamma} \langle\tau_{\text{ph}}\rangle_{\gamma\beta}
\label{eq.rta2},
\end{equation}
where $\langle\tau_{\text{ph}}\rangle_{\gamma\beta}$ is defined by the matrix equation:
\begin{eqnarray}
\langle\tau_{\text{ph}}\rangle &=& 
\left[ \left\langle\!\frac{1}{\tau_{\text{ph}}^{\text{(3ph)}}}\!\right\rangle
+\left\langle\!\frac{1}{\tau_{\text{ph}}^{\text{(4ph)}}}\!\right\rangle \right. \nonumber\\
&& \left.+\left\langle\!\frac{1}{\tau_{\text{ph}}^{\text{(elph)}}}\!\right\rangle
+\left\langle\!\frac{1}{\tau_{\text{ph}}^{\text{(iso)}}}
\!\right\rangle \right]^{-1}.
\end{eqnarray}
Upon replacing Eqs.~(S5) inside (S4), and using (S3), one correctly recovers the RTA condutivity given by Eq.~(S1). We report the isotropically-averaged average scattering rates thus determined in Table~\ref{t:scattcomp}.

\section{Theoretical analysis of thermal conductivity in $\delta$-TaN}\label{app:delta}
Figure~\ref{f:delta} summarizes the calculated thermal-transport, phonon, and electronic properties of $\delta$-TaN. As shown in Fig.~\ref{f:delta}(a), heat conduction in $\delta$-TaN is dominated by electrons, as in conventional metals. At room temperature, we obtain $\kappa_{\mathrm{el}}=64$~W\,m$^{-1}$\,K$^{-1}$ and $\kappa_{\mathrm{ph}}=11$~W\,m$^{-1}$\,K$^{-1}$, resulting in a total thermal conductivity of $\kappa_{\mathrm{tot}}=75$~W\,m$^{-1}$\,K$^{-1}$. This value is substantially smaller than those of the $\varepsilon$ and $\theta$ phases. Nevertheless, the electronic thermal conductivity is slightly higher than that of $\varepsilon$-TaN, consistent with the higher Fermi velocity reported in Table~\ref{t:fv}.

Fig.~\ref{f:delta}(b) shows the decomposition of $\kappa_\text{ph}$ into scattering channels. The dominant scattering mechanism is electron-phonon scattering, as shown in Tab.~\ref{t:scattcomp}. Both electron-phonon and four-phonon scattering are stronger in $\delta$-TaN than in the other two phases.

The harmonic phonon dispersion of $\delta$-TaN exhibits soft modes and a spurious lattice instability, consistent with previous reports~\cite{Isaev2007,Yan2020}. The inclusion of anharmonic corrections stabilizes these modes, as illustrated in Fig.~\ref{f:delta}(c). The anharmonically renormalized dispersion also exhibits an acoustic-optical gap similar to that of $\theta$-TaN. However, the speed of sound, $v_{\mathrm{s}}=4.8\times10^{3}$~m/s, is smaller than those of the other TaN phases considered here.

Figs.~\ref{f:delta}(d)-(f) show the electronic band structure, density of states, and Fermi surface of $\delta$-TaN, respectively. Although its Fermi velocity remains high, $v_{\mathrm{F}}=1.1\times10^{6}$~m/s, $\delta$-TaN has a larger DOS at the Fermi level than $\varepsilon$-TaN. Together with its relatively complex Fermi surface, this large density of state provides substantial phase space for electron-phonon scattering, resulting in strong phonon damping and reduced phonon relaxation times.

\bibliography{ref}

\end{document}